\documentclass[aps,prl,notitlepage,twocolumn,longbibliography,superscriptaddress]	{revtex4-1}

\usepackage[colorlinks=true,citecolor=blue]{hyperref}
\usepackage{amsmath,mathtools,latexsym,amssymb,bm,color,graphicx,multirow,array,wrapfig,scalerel,datetime,enumerate}
\usepackage{tabularx}
\newcolumntype{C}{>{\centering\arraybackslash}X}%

\usepackage{subcaption}
\usepackage{ragged2e}
\DeclareCaptionJustification{justified}{\justifying}
\newcommand{\dd}{\mathrm{d}}

\newcommand{\s}{\sigma}
\newcommand{\om}{\omega}

\newcommand{\ra}{\rightarrow}

\newcommand{\Z}{\mathbb{Z}}
\newcommand{\Zf}{\mathbb{Z}_2^f}
\newcommand{\U}{\mathrm{U(1)}}
\newcommand{\Uf}{\mathrm{U(1)}_f}

\newcommand{\eq}[1]{Eq.~(\ref{#1})}
\newcommand{\eqs}[2]{Eqs.~(\ref{#1}) and (\ref{#2})}
\newcommand{\fig}[1]{Fig.~\ref{#1}}
\newcommand{\tab}[1]{Table~\ref{#1}}
\graphicspath{{figures/}}

\begin{document}

\title{Exactly solvable lattice models for interacting electronic insulators in two dimensions}

\author{Qing-Rui Wang}
\affiliation{Yau Mathematical Sciences Center, Tsinghua University, Haidian, Beijing 100084, China}
\affiliation{Yanqi Lake Beijing Institute of Mathematical Sciences and Applications, Huairou, Beijing 101408, China}
\author{Yang Qi}
\affiliation{Center for Field Theory and Particle Physics, Department of Physics, Fudan University, Shanghai 200433, China}
\affiliation{State Key Laboratory of Surface Physics, Fudan University, Shanghai 200433, China}
\author{Chen Fang}
\affiliation{Beijing National Laboratory for Condensed Matter Physics and Institute of Physics, Chinese Academy of Sciences, Beijing 100190, China}
\author{Meng Cheng}
\affiliation{Department of Physics, Yale University, New Haven, CT 06511-8499, USA}
\author{Zheng-Cheng Gu}
\affiliation{Department of Physics, The Chinese University of Hong Kong, Shatin, New Territories, Hong Kong, China}


\begin{abstract}
In the past decade, tremendous efforts have been made towards understanding fermionic symmetry protected topological (FSPT) phases in interacting systems. Nevertheless, for systems with continuum symmetry, e.g., electronic insulators, it is still unclear how to construct an exactly solvable model with a finite dimensional Hilbert space in general. In this paper, we give a lattice model construction and classification for 2D interacting electronic insulators. Based on the physical picture of $\mathrm{U(1)}_f$-charge decorations, we illustrate the key idea by considering the well known 2D interacting topological insulator. Then we generalize our construction to an arbitrary 2D interacting electronic insulator with symmetry $G_f=\mathrm{U(1)}_f \rtimes_{\rho_1,\omega_2} G$, where $\mathrm{U(1)}_f$ is the charge conservation symmetry and $\rho_1, \omega_2$ are additional data which fully characterize the group structure of $G_f$. Finally we study more examples, including the full interacting classification of 2D crystalline topological insulators.
\end{abstract}

\maketitle



\emph{Introduction.---}
In recent years, remarkable progress has been made in the theoretical understanding of gapped phases in quantum many-body systems, in particular for fermionic symmetry-protected topological (FSPT) phases~\cite{fidkowski10,fidkowski11,chen11b,GuWen2014,
GuLevin2014,Kapustin2015,wangc-science,Senthil2013,
Lukasz2013,XLQi2013,XieChen2014,metlitski15,ChongWang2014,Witten,
Freed16,wanggu16,Tarantino2016,Kapustin2017,fbraiding,WG18,WG20} 
, which include topological band insulators as the most familiar example~\cite{ZhangRMP,KaneRMP}. Exactly solvable lattice Hamiltonians, whose ground states are fixed-point wavefunctions, have played a vital role in these development, which often serve as proof-of-principle models for the existence of interacting topological phases and facilitate extraction of universal physical properties to characterize the topological order. They can often be turned into exact tensor network states, offering a convenient starting point for the study of more realistic systems. However, known constructions of SPT phases typically feature local Hilbert space isomorphic to the protecting symmetry group, which becomes problematic if the symmetry is continuous. 
To date, no systematic exactly-solvable constructions are available for generic electronic insulators, except for a couple of isolated examples. In this paper, we generalize the decorated domain wall construction of interacting FSPT with finite total symmetry group $G_f$ into interacting electronic insulators involving $\Uf$ charge conservation symmetry. As a simple application, we will derive the full interacting classification of 2D crystalline topological insulators~\cite{Metlitski2019,Son2019}. Our method can also be applied to systems with other continuum symmetry such as $\mathrm{SU(2)}$ spin rotational symmetry. 

\emph{2D interacting topological insulator from $\Uf$-charge decorations.---}
We begin with a concrete example of 2D FSPT state protected by $G_f
=(\Uf\rtimes \Z_4^T)/\Z_2$. 
It is the well known 
topological insulator with $\Uf$ charge conservation and time reversal symmetries, where fermions transform as Kramers doublets under time reversal. 

\begin{figure}[t]
\centering
$\vcenter{\hbox{\includegraphics[scale=1]{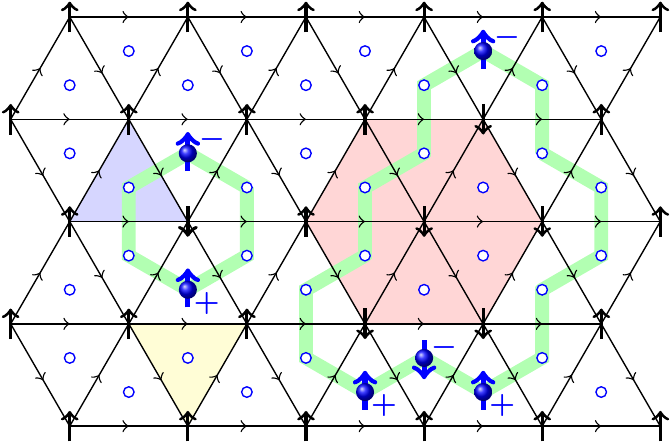}}}$
\caption{Fermionic $\Uf$ charge decoration.
Fermions with $\Uf$ charge $+1$ and $-1$ (blue dots) are decorated at the minimum and maximum points of the domain walls (green lines), respectively.
The spin of the fermion (blue arrow) depends on the bosonic spin (black arrow) at the left vertex $\s_i$ of the corresponding triangle.
The terms $P_\bigtriangleup$, $P_\bigtriangledown$ and $A_s$ of the Hamiltonian are associated with triangles illustrated by blue, yellow and red colors, respectively.}
\label{fig:H}
\end{figure}

Let us consider a triangular lattice shown in \fig{fig:H}. On each vertex $i$, there is a bosonic Ising spin $\s_i=\uparrow\!/\!\downarrow=\pm 1$. 
At the center of each triangle $\langle ijk\rangle$, there is a spin-1/2 fermionic degrees of freedom $c_{ijk}^\s$ ($\s=\, \uparrow\!/\!\downarrow$). While the bosonic spin $\s_i$ does not carry $\Uf$ charge, the $\Uf$ charge of the fermion $c_{ijk}^\s$ is chosen to be $+1$ ($-1$) if $\langle ijk\rangle$ is an up-pointing triangle $\bigtriangleup$ (a down-pointing triangle $\bigtriangledown$). On the other hand, the time reversal symmetry flips the bosonic spin $\s_i$ between $\uparrow$ and $\downarrow$, and transforms the spin-1/2 fermion as $c_{ijk}^\uparrow \rightarrow c_{ijk}^\downarrow$ and $c_{ijk}^\downarrow \rightarrow -c_{ijk}^\uparrow$. 

The fixed-point wavefunction is obtained by decorating fermionic $\Uf$ charges to the symmetry domain walls of $\{\s_i\}$~\footnote{A simpler bosonic U(1) charge decoration can be found in Refs.~\onlinecite{horinouchi2020solvable} and \onlinecite{wang2021exactly}.}. To be more specific, let us consider the domain wall configurations of a single triangle $\langle ijk\rangle$. There are in total $2^3=8$ different spin (black arrow) configurations or 4 domain wall (green line) configurations, for example, in an up-pointing triangle:
\begin{equation}\label{triangle}
\begin{split}
\vcenter{\hbox{\includegraphics[scale=1]{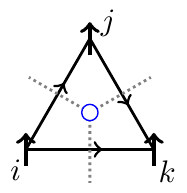}}}
\vcenter{\hbox{\includegraphics[scale=1]{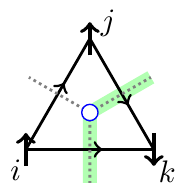}}}
\vcenter{\hbox{\includegraphics[scale=1]{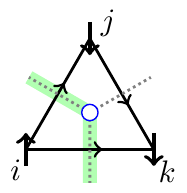}}}
\vcenter{\hbox{\includegraphics[scale=1]{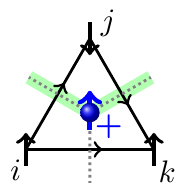}}}
\\
\vcenter{\hbox{\includegraphics[scale=1]{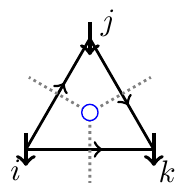}}}
\vcenter{\hbox{\includegraphics[scale=1]{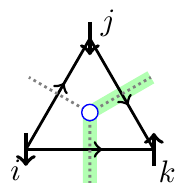}}}
\vcenter{\hbox{\includegraphics[scale=1]{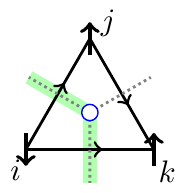}}}
\vcenter{\hbox{\includegraphics[scale=1]{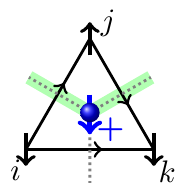}}}
\end{split}
\end{equation}
If the configuration satisfies $\s_i=-\s_j=\s_k$ (see the two rightmost figures above), a fermion $c_{ijk}^{\s_i}$ with spin $\s_i$ and $\Uf$ charge $+1$ ($-1$) will be decorated at the center when the triangle $\langle ijk\rangle$ is up-pointing (down-pointing). An explicit example of the decorations can be found in \fig{fig:H}.
The fixed-point wavefunction is a superposition of all possible bosonic spin configurations decorated with fermionic $\Uf$ charges using the rules above:
\begin{equation}\label{TIwf}
|\Psi\rangle = \sum_{\text{all conf.}} \Psi\left(
\ 
\vcenter{\hbox{\includegraphics[scale=1]{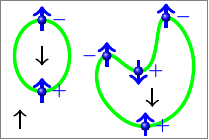}}}
\ 
\right) \stretchleftright{\Bigg|}{\ 
\vcenter{\hbox{\includegraphics[scale=1]{figures/Fig_wf_1}}}
\ }{\Big\rangle}.
\end{equation}
By solving the consistency conditions (symmetry condition and twisted super-cocycle equation), we will show later that the coefficient $\Psi(c)$ for each configuration $c$ is always $\pm 1$ 
depending on the order of the decorated fermions.

The above $\Uf$-charge decoration is compatible with the symmetry $G_f=\Uf\rtimes_{\rho_1,\om_2} \Z_2^T$. For each domain wall loop, the numbers of minimum and maximum points are the same. 
Therefore, the total $\Uf$ charge of the decorated configuration is always zero. On the other hand, the time reversal symmetry flips all the bosonic and fermionic spins in the configuration. 
By choosing the coefficient $\Psi(c)$ appropriately, one can make the ground state $|\Psi\rangle$ time reversal invariant.

\emph{Commuting-projector Hamiltonian and edge state.---}
As a fixed-point wavefunction, the $\Uf$-charge decorated state \eq{TIwf} is the ground state of an exactly-solvable commuting-projector Hamiltonian with finite-dimensional local Hilbert spaces (see \fig{fig:H}):
\begin{equation}
H=-\sum_\bigtriangleup P_\bigtriangleup - \sum_\bigtriangledown P_\bigtriangledown - \sum_{\text{site }s} \frac{1+A_s}{2} \prod_{\bigtriangleup}P_{\bigtriangleup} \prod_{\bigtriangledown}P_{\bigtriangledown}. \label{H}
\end{equation}
The triangle terms $P_\bigtriangleup$ and $P_\bigtriangledown$ are projections enforcing the decoration rules such as \eq{triangle} for each triangle. The operator $A_s$ in the last term flips the bosonic spin at site $s$, and changes the fermionic $\Uf$ charge decorations accordingly for the six surrounding triangles. We present more details of the Hamiltonian in the Supplemental Material. In the literature, there are other constructions for the interacting topological insulator. Compared to the method of decorating multiple Majorana chains~ \cite{Metlitski2019,Son2019}, the state \eq{TIwf} we constructed is much simpler and can be systematically generalized to other symmetry group $G_f$, which we will describe later. 

The state \eq{TIwf} is the interacting counterpart of the free-fermion topological insulator with charge conservation and time-reversal symmetries. They share the same nontrivial gapped, symmetry-breaking edge state. In fact, we can consider a position-dependent Zeeman field on the boundary, such that there are two edge spin domain walls, whose local profile are related to each other via time-reversal symmetry. Due to the $\Uf$ charge conservation of the domain wall loop, these two edge domain wall should have total $\Uf$ charge $\pm 1$. If the edge is particle-hole symmetric, each domain wall will have half $\Uf$ charge (see Supplemental Material for formal derivation).

\emph{Symmetries of interacting electronic insulators.---}
Before generalizing the above constructions to other systems, we first need to introduce some notations and definitions about the symmetry group $G_f$
. For insulators, there is a $\Uf$ charge conservation symmetry. The element of this group is $U_\theta=e^{i\theta Q}$, where 
$Q$ is the $\Uf$ charge operator. As the fermion parity operator is the order-2 element $U_\pi$ in this group, we will denote the charge conservation symmetry by $\Uf$ with a subscript $f$. The action of $U_\theta$ on a bosonic/fermionic annihilation operator with $\Uf$ charge $q$ is $U_\theta c_j^{\sigma,q} U_\theta^\dagger = e^{- iq\theta} c_j^{\sigma,q}$, where $j$ is the lattice site and $\sigma$ is the combination of other indices such as orbital and spin, \emph{etc}. As $\Uf$ charge symmetry is always a normal subgroup of the total symmetry $G_f$ for electronic insulators,
we have the following short exact sequence:
\begin{align}
1 \ra \Uf \ra G_f \ra G \ra 1,
\end{align}
where $G:=G_f/\Uf$ is the quotient group. In this paper, we assume that $G$ is a finite group.

Conversely, given $\Uf$ and $G$, we can recover the group $G_f=\Uf \rtimes_{\rho_1,\om_2} G$ by using two ingredients $\rho_1$ and $\om_2$. 
The 1-cocycle $\rho_1\in H^1(G,\Z_2)$ is a homomorphism from $G$ to $\operatorname{Aut}(\Uf)=\Z_2$. It implements the charge conjugation action of $G$ on $U_\theta=e^{i\theta Q}\in\Uf$ as
\begin{align}\label{rho1}
g \times U_\theta \times g^{-1} = (U_\theta)^{(-1)^{\rho_1(g)}} = U_{(-1)^{\rho_1(g)}\theta}.
\end{align}
The second ingredient $\om_2$ is related to the extension of $G$. As a set, $G_f$ is the same as $\Uf\times G$,  so the elements of $G_f$ can be parametrized as $(U_\theta, g)$. But the multiplication in $G_f$ reads
\begin{align}
(1,g)\times (1,h) = \left(U_{2\pi \om_2(g,h)},gh\right) \in G_f,
\end{align}
where $\om_2(g,h)\in \mathbb{R}/\mathbb{Z}\simeq\Uf$ is a phase associated with $g,h\in G$. 
The associativity condition of $G_f$ implies that $\om_2$ is a 2-cocycle in $H^2_{\rho_1}(G,\Uf)$~\footnote{Since cohomologous $\om_2$'s will give isomorphic $G_f$, we also have to mod out the 2-coboundaries.}, where the subscript $\rho_1$ indicates the $G$-action on the coefficient $\Uf$.


The two cocycles $\rho_1$ and $\om_2$ fully characterize the group structure of $G_f=\Uf \rtimes_{\rho_1,\om_2} G$, but the action of the group $G$ or $G_f$ on the 
wavefunctions
is still not full determined yet. When there is an anti-unitary symmetry in $G$, we should also introduce a third ingredient $s_1$ to specify its action on the wavefunctions with $i \rightarrow -i$:
\begin{align}\label{s1}
s_1(g)=
\begin{cases}
0, & \text{if $g$ is unitary,}\\
1, & \text{if $g$ is anti-unitary}.
\end{cases}
\end{align}
Apparently, $s_1$ is also a 1-cocycle in $H^1(G,\Z_2)$.

In general, the 1-cocycles $s_1$ and $\rho_1$ are not the same. Combining Eqs.~(\ref{rho1}) and (\ref{s1}), the $G$ action on the charge operator $Q$ in $U_\theta=e^{i\theta Q}\in \Uf$ should be
\begin{align}\label{Q}
g \times Q \times g^{-1} = (-1)^{\rho_1(g)+s_1(g)} Q.
\end{align}
So the $\Uf$ charges change sign under the $g$ action if and only if $\rho_1(g)$ and $s_1(g)$ are different.


\emph{Generalization to symmetry $G_f=\Uf \rtimes_{\rho_1,\om_2} G$.}---
Now we want to generalize the construction of $\Uf$ charge decoration to arbitrary 2D interacting electronic insulators protected by $G_f=\Uf \rtimes_{\rho_1,\om_2} G$. 
The degrees of freedom (d.o.f.) of our lattice model is as follows. We first triangulate the 2D spacial manifold with a branching structure. On each vertex $i$, we put a $|G|$-level spin Hilbert space spanned by $|g_i\rangle$ ($g_i\in G$). At the center of each triangle $\langle ijk \rangle$, we put a Hilbert space spanned by bosons/fermions $c_{ijk}^{\sigma,q}$ ($\sigma\in G$, $q\in \Z$, $|q|<\Lambda$). Here $q$ is the $\Uf$ charge of the boson/fermion, and $\Lambda$ is a finite positive integer depending on $G$ \footnote{$\Lambda$ is the biggest number of $|n_2(g_i,g_j,g_k)|$ for all possible $g_i,g_j,g_k\in G$ and 2-cocycles $n_2\in H^2_{\rho_1+s_1}(G,\Z)$.}.
We choose the d.o.f. $c_{ijk}^{\sigma,q}$ to be a fermion (boson) if $q$ is odd (even) \footnote{One can think of the $q=\pm 1$ fermion to be the fundamental $\Uf$ charges. All other $q$ charges are combinations of several fundamental charges. Thus odd (even) $q$ corresponds to fermion (boson).}. So the (anti-)commutation relation reads
\begin{equation}
c_{ijk}^{\s,q} (c_{i'j'k'}^{\s',q'})^\dagger - (-1)^{qq'} (c_{i'j'k'}^{\s',q'})^\dagger c_{ijk}^{\s,q} = \delta_{ijk,i'j'k'} \delta_{\s\s'} \delta_{qq'}.
\end{equation}
Under the symmetries $U_\theta\in \Uf$ and $g\in G$, these d.o.f. transform as:
\begin{eqnarray}
&&U_\theta |g_i\rangle = |g_i\rangle,\ 
U(g) |g_i\rangle = |gg_i\rangle,\ 
U_\theta c_{ijk}^{\s,q} U_\theta^\dagger = e^{- i q\theta} c_{ijk}^{\s,q},\nonumber\\ 
&&U(g) c_{ijk}^{\s,q} U(g)^\dagger \nonumber\\ &&= e^{-2\pi i \om_2(g,\s) (-1)^{\rho_1(g)+s_1(g)}q} c_{ijk}^{g\s,(-1)^{\rho_1(g)+s_1(g)}q}.\label{Ugc}
\end{eqnarray}
In this way, both the bosonic and fermionic d.o.f. support linear representations of the total symmetry group $G_f$ (see Supplemental Material for a proof). 

To obtain a 2D $G_f$-FSPT state, we can decorate $\Uf$ charges to the domain wall junctions of $G$. After proliferating $G$ domain walls, we will obtain a symmetric gapped FSPT state protected by symmetry $G_f$. Schematically, the wavefunction would have the form
\begin{equation}
\label{wf}
|\Psi\rangle = \sum_{\text{all conf.}} \Psi\left(
\ 
\vcenter{\hbox{\includegraphics[scale=1]{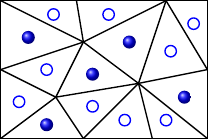}}}
\ 
\right) \stretchleftright{\Bigg|}{\ 
\vcenter{\hbox{\includegraphics[scale=1]{figures/Fig_wf2_1}}}
\ }{\Big\rangle}, \nonumber
\end{equation}
where the blue dots are the decorated $\Uf$ charges similar to \eq{TIwf}.
Now we try to decorate the $\Uf$ charges $c_{ijk}^{\sigma,q}$ to the domain wall junctions (triangle centers) of $G$. The decoration is specified by an integral charge function $n_2(g_i,g_j,g_k)\in \Z$. For a triangle $\langle ijk\rangle$ with orientation $r_{ijk}=\pm 1$ and vertex spin labels $e,g_0^{-1}g_1,g_0^{-1}g_2\in G$, we decorate the $\Uf$ charge $c_{ijk}^{e,r_{ijk}n_2(e,g_0^{-1}g_1,g_0^{-1}g_2)}$ at the center. All other charges $c_{ijk}^{\sigma,q}$ of this triangle with $\s\ne e$ or $q\ne r_{ijk}n_2(e,g_0^{-1}g_1,g_0^{-1}g_2)$ remain empty or in the vacuum state. From this standard triangle decoration, we can obtain the decoration for arbitrary triangle under the action of $U(g_0)$:
\begin{equation}\label{fig:012}
\vcenter{\hbox{\includegraphics[scale=1]{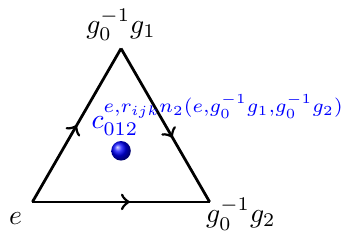}}}
\xrightarrow{U(g_0)}
\vcenter{\hbox{\includegraphics[scale=1]{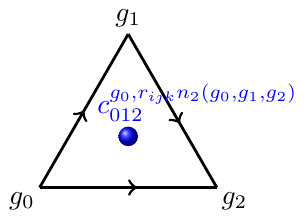}}}
. 
\end{equation}
To be consistent with the symmetry transformation \eq{Ugc}, the function $n_2$ should satisfy:
\begin{equation}\label{n2symm}
n_2(g_0,g_1,g_2) = (-1)^{\rho_1(g_0)+s_1(g_0)} n_2(e,g_0^{-1}g_1,g_0^{-1}g_2). 
\end{equation}
So $n_2$ is a 2-cochain in $C_{\rho_1+s_1}^2(G,\Z)$ with a $G$-action on the integral charges indicated by the subscript $\rho_1+s_1$. This nontrivial action can be traced back to \eq{Q}.


\emph{$\Uf$-symmetric fermionic $F$ moves.---}
To make the wavefunction \eq{wf} well-defined, we have to check several consistency conditions. The easiest way is to consider wavefunctions on different triangulations of the spacial manifold. They are related to each other by elementary local changes called Pachner moves ($F$ moves). 
Since we want the wavefunction to be $G_f$-symmetric, the $F$ moves should respect the symmetry. So we have the following commuting square:
\begin{align}\label{2D:FUg}
\vcenter{\hbox{\includegraphics[scale=1]{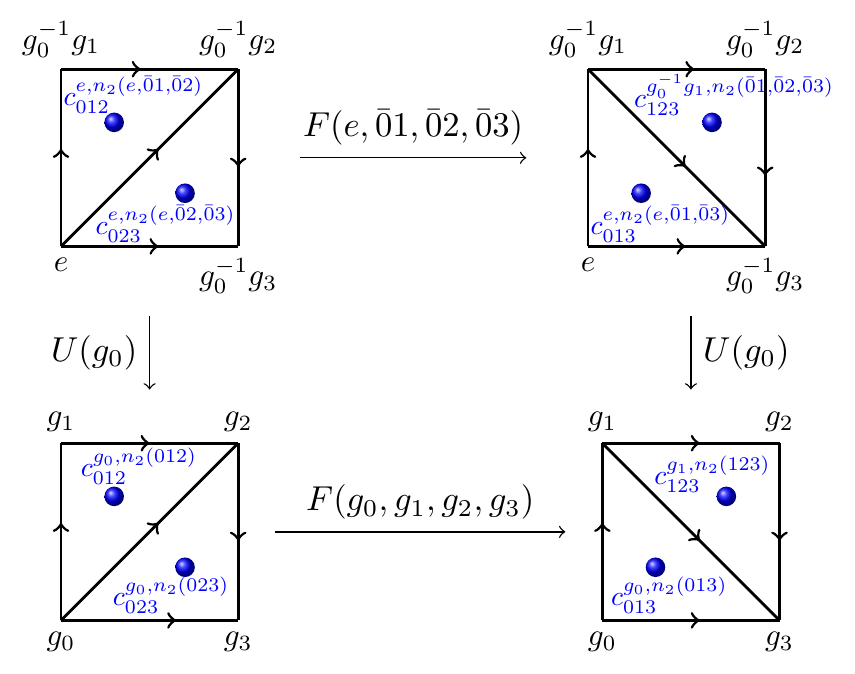}}}
.
\end{align}
Given the standard $F$ move with the first vertex labelled by $e\in G$, we can use the above commuting diagram to derive the non-standard one with generic $g_0\in G$. They have the following explicit expressions:
\begin{widetext}
\begin{align}\label{Fe}
F(e,\bar 01,\bar 02,\bar 03)
&:=
\nu_3(e,\bar 01,\bar 02,\bar 03)
\big(c_{012}^{e,n_2(e,\bar 01,\bar 02)}\big)^{\dagger}
\big(c_{023}^{e,n_2(e,\bar 02,\bar 03)}\big)^{\dagger}
c_{013}^{e,n_2(e,\bar 01,\bar 03)}
c_{123}^{g_0^{-1}g_1,n_2(\bar 01,\bar 02,\bar 03)},\\\nonumber
F(g_0,g_1,g_2,g_3)
&=
U(g_0) F(e,g_0^{-1}g_1,g_0^{-1}g_2,g_0^{-1}g_3) U(g_0)^{-1}
\\\label{Fg0}
&:= \nu_3(g_0,g_1,g_2,g_3)
\big(c_{012}^{g_0,n_2(012)}\big)^{\dagger}
\big(c_{023}^{g_0,n_2(023)}\big)^{\dagger}
c_{013}^{g_0,n_2(013)}
c_{123}^{g_1,n_2(123)},
\end{align}
\end{widetext}
where we use abbreviations $\bar ij$ for $g_i^{-1}g_j$ and $n_2(ijk)$ for $n_2(g_i,g_j,g_k)$. We also set $r_{ijk}=1$ for all the triangles shown above.
From the $U(g_0)$-action on the complex numbers and bosonic/fermionic $\Uf$ charges in \eq{Ugc}, the $F$ move coefficient $\nu_3\in C_{s_1}^3(G,\U)$ has the symmetry condition
\begin{align}\nonumber\label{symm_sign}
\nu_3(g_0,g_1,g_2,g_3) &= \left[\nu_3(e,g_0^{-1}g_1,g_0^{-1}g_2,g_0^{-1}g_3)\right]^{1-2s_1(g_0)}\\
&\quad\times e^{-2\pi i \om_2(g_0,g_0^{-1}g_1) n_2(g_1,g_2,g_3)},
\end{align}
Here we use the normalization condition $\om_2(g_0,e)=0$.

Besides the $G$ symmetry, the $F$ should also preserve the $\Uf$ charges. By counting the $\Uf$ charges on the two sides of the $F$ move \eq{Fe}, we have the integer equation:
\begin{eqnarray}
\label{dn2}
&& 
(\dd_{\rho_1+s_1} n_2)(g_1,g_2,g_3)\\\nonumber
&=& (-1)^{\rho_1(g_1)+s_1(g_1)}n_2(g_2,g_3) - n_2(g_1g_2,g_3)\\\nonumber
&+& n_2(g_1,g_2g_3) - n_2(g_1,g_2)= 0,
\end{eqnarray}
where we define the inhomogeneous cochain $n_2(g_1,g_2):=n_2(e,g_1,g_1g_2)$ to be the homogeneous one with the first argument being $e\in G$. One can also show that adding coboundaries to $n_2$ can be gauged away by symmetric local unitaries. Therefore, $n_2$ is in fact a 2-cocycle in $H_{\rho_1+s_1}^2(G,\Z)$. Here, we use the subscript $\rho_1+s_1$ to indicate the possibly nontrivial $G$-action on the $\Uf$ charge appearing in the first term of the second line of \eq{dn2}. This action originates from \eqs{Q}{n2symm}.

\emph{Twisted super-cocycle equation.---}
Given two triangulations of the spacial manifold, there are possibly many different sequences of $F$ moves connecting them. Since the initial and the final states are fixed, we should have the same result from different sequences. The smallest loop among these sequences is the twisted version of super-cocycle equation~\cite{GuWen2014}. 

Let us choose the label of the first vertex to be $e\in G$. In this way, the standard super-cocycle equation reads
\begin{align}\nonumber\label{dF}
&\quad F(e,\bar 01,\bar 02,\bar 03) \cdot F(e,\bar 01,\bar 03,\bar 04) \cdot F(\bar 01,\bar 02,\bar 03,\bar 04)\\
&= F(e,\bar 02,\bar 03,\bar 04) \cdot F(e,\bar 01,\bar 02,\bar 04).
\end{align}
The non-standard ones are automatically satisfied by simply a symmetry action $U(g_0)$. Using the symmetry condition $F(\bar 01,\bar 02,\bar 03,\bar 04)=U(\bar 01)F(e,\bar 12,\bar 13,\bar 14)U(\bar 01)^\dagger$ from \eq{Fg0}, we can convert the above equation to a formula that only involves the standard $F$ moves \eq{Fe}. After eliminating all the $c_{ijk}^{\s,q}$ operators, the final result is a twisted cocycle equation for the $\nu_3$ as
\begin{align}\label{dnu3}
\dd_{s_1} \nu_3 &= e^{2\pi i(\omega_2 \smile n_2 + \frac{1}{2}n_2\smile n_2) }.
\end{align}
Here, the differential $\dd_{s_1}$ of the inhomogeneous cochain $\nu_3(g_1,g_2,g_3) := \nu_3(e,g_1,g_1g_2,g_1g_2g_3)$ is defined as
\begin{align}
&\quad\ (\dd_{s_1} \nu_3)(g_1,g_2,g_3,g_4)\\\nonumber
&=
\frac{\nu_3(g_2,g_3,g_4)^{1-2s_1(g_1)} \nu_3(g_1,g_2g_3,g_4) \nu_3(g_1,g_2,g_3)}{\nu_3(g_1g_2,g_3,g_4) \nu_3(g_1,g_2,g_3g_4)},
\end{align}
and the first cup product on the right-hand side of \eq{dnu3} reads
\begin{align}\nonumber
&\quad\ (\omega_2 \smile n_2)(g_1,g_2,g_3,g_4)\\
&=
\om_2(g_1,g_2) (-1)^{\rho_1(g_1g_2)+s_1(g_1g_2)} n_2(g_3,g_4).
\end{align}
It has a simpler expression $(\omega_2 \smile n_2)(e,g_1,g_2,g_3,g_4) = \om_2(e,g_1,g_2) n_2(g_2,g_3,g_4)$ in the homogeneous notation, where the $G$-action sign $(-1)^{\rho_1+s_1}$ is absorbed in $n_2(g_2,g_3,g_4)$. The second cup product $(-1)^{n_2\smile n_2}$ has a similar expression and comes from the reordering of the $c_{ijk}^{\s,q}$ operators when they are fermions.


Using the solutions $(n_2,\nu_3)$ of the obstruction equations \eqs{dn2}{dnu3}, we can construct a $G_f$-symmetric wavefunction \eq{wf} by decorating $\Uf$ charges. It can be shown that the decoration data $(n_2,\nu_3)$ of the same cohomology class would give us equivalent wavefunctions related by fermionic symmetric local unitary transformations. Moreover, as discussed in the Supplemental Material, $\nu_3$ and $\nu_3 e^{2\pi i \om_2\smile n_1}$ with $n_1\in H^1_{\rho_1+s_1}(G,\Z)$ are also equivalent.
Therefore, the final classification data of interacting electronic insulators are $n_2$ and $\nu_3$, which are elements in $H_{\rho_1+s_1}^2(G,\Z)$ and $C_{s_1}^3(G,\U)/B_{s_1}^3(G,\U)/\Gamma^3$, where $\Gamma^3$ is the trivialization subgroup due to the 1D anomalous SPT states \cite{WQG2018}. 

\emph{More Examples.---}
Let us consider some simple examples of $G_f$-FSPT with charge conservation symmetry.

(1) $G_f=\Uf\times \Z_2$. In this case, we have $G=\Z_2$ and $\rho_1=s_1=\om_2=0$. It can be shown easily that the nontrivial fermion decoration $n_2\in H^2(\Z_2,\Z)$ is obstruction-free. After gauging $G$ and considering only the $\Z_2^f$ subgroup of $\Uf$, the state is identical to the fermionic toric code~\cite{GWW}. With a nontrivial BSPT protected by $G$ only, the full classification of $G_f$-FSPT is $\Z_4$. In fact, the root state of this $\Z_4$ is the $\nu=2$ state of the $\Z_8$ classification of $G_f=\Zf\times\Z_2$ FSPT~\cite{GuLevin2014}.

(2) $G_f=\Uf\rtimes \Z_2^T$. Now $\rho_1=s_1$ is nontrivial and $\om_2$ is trivial. One can show that the $\Uf$-charge decoration $n_2$ is obstructed. There is also no BSPT state. So there is only a trivial $G_f$-FSPT state.

(3)By applying the fermionic crystalline equivalence~\cite{reduction,correspondence,rotation,resolution,dihedral,JH2020} where a mirror reflection symmetry action should be mapped onto a time-reversal symmetry action, and that spinless (spin-1/2) fermionic systems should be mapped into spin-1/2 (spinless) fermionic systems, we can also derive the complete interacting classification of 2D crystalline topological insulators. In Supplementary Material, we list the classification results for all 17 wall paper groups. 





\emph{Discussion and conclusion.---}
In this paper, we construct and classify interacting electronic insulators in two spacial dimensions with arbitrary symmetry group $G_f=\Uf\rtimes_{\rho_1,\om_2}G$. The construction is obtained by decorating $\Uf$ charges to the $G$ symmetry domain wall junctions. This decoration is specified by a 2-cocycle $n_2\in H_{\rho_1+s_1}^2(G,\Z)$. The second piece of  classification data $\nu_3\in C_{s_1}^3(G,\U)/B_{s_1}^3(G,\U)/\Gamma^3$ is the wavefunction coefficient satisfying the super-cocycle equation (\ref{dnu3}). As an explicit example, we construct the fixed-point wavefunction and commuting-projector Hamiltonian of topological insulator with charge conservation and time-reversal symmetries. By applying the crystalline equivalence principle, we also derive the complete interacting classification of 2D crystalline topological insulators. Apparently, our classification data can also classify interacting electronic insulators with both internal and space group symmetry.

Finally,  we stress that our constructions and classification scheme can be easily generalized to other continuous groups by decorating the corresponding continuous-symmetry-protected states to discrete-symmetry domain walls. It can be also generalized from two dimensions to higher dimensions, though the corresponding obstruction functions could become more complicated.

\emph{Acknowledgements.---}
Z.C.G. is supported by Direct Grant No. 4053462 from The Chinese University of Hong Kong and funding from Hong Kong's Research Grants Council (GRF No.14306420, ANR/RGC Joint Research Scheme No. A-CUHK402/18). Y.Q. is supported by the National Natural Science Foundation of China (Grant No. 11874115). M.C. acknowledges support from NSF under award number DMR-1846109.


%

\newpage
\begin{widetext}

\appendix
\begin{center}
\large\bf Supplemental Material
\end{center}

\section{A. Ground-state wavefunction for 2D interacting TI}

In this section of the Supplemental Material, we will derive some consistency conditions for the coefficient $\Psi(c)$ of the ground state \eq{TIwf} of the 2D interacting topological insulator protected by $\Uf$ charge and time reversal symmetries.

Using the notion of fermionic symmetric local unitary moves, we can relate the coefficient of one configuration to that of another. There are several basic local moves. Using a sequence of basic local moves, we can obtain the coefficient $\Psi(c)$ for each configuration $c$ from the vacuum configuration.

\subsection{B.1. Basic local moves}

Instead of directly deriving the coefficient $\Psi(c)$ for each configuration, we first try to understand the relation between $\Psi(c)$ and $\Psi(c')$ for two different configurations $c$ and $c'$. In general, we can use a sequence of basic moves to deform the configuration $c$ to $c'$. Each basic move only changes a local patch of the configuration. If the coefficient changes of the basic moves are known, we can use them to obtain $\Psi(c)$ of arbitrary configuration $c$ from the special configuration $c_0$ with no domain walls and fermions.

(1) Domain wall shape changing.
If we only deform the domain wall shape without changing the numbers of minimum and maximum points, there is no creation or annihilation of fermions. So the coefficient of the fixed-point wavefunction remains the same:
\begin{align}\label{shape}
\Psi\left(
\ \vcenter{\hbox{\includegraphics[scale=1]{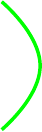}}}\ 
\right)
=
\Psi\left(
\ \vcenter{\hbox{\includegraphics[scale=1]{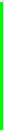}}}\ 
\right).
\end{align}

(2) Creation and annihilation of local domain wall loop. If a domain wall loop is created in a configuration, there should be two fermions decorated at the minimum and maximum point of the loop. So the wavefunction coefficient will be changed as
\begin{align}\label{O}
\Psi\left(
\ \vcenter{\hbox{\includegraphics[scale=1]{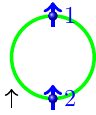}}}\ 
\right)
&=
c_{1\uparrow}^\dagger c_{2\uparrow}^\dagger
\Psi\left(
\ \vcenter{\hbox{\includegraphics[scale=1]{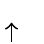}}}\ 
\right).
\end{align}
Here, the labels $1$ and $2$ of the left configuration indicate the creation order of the fermions. The up arrow at the bottom-left corner is the spin of the bosonic spins outside the domain wall loops. And the right configuration is the one with only up bosonic spins and no domain walls in this local patch.

Under the time reversal symmetry action, bosonic and fermionc spins are flipped with plus or minus signs ($c_{ijk}^\downarrow \rightarrow -c_{ijk}^\uparrow$). So the symmetry partner of \eq{O} is
\begin{align}\label{Odown}
\Psi\left(
\ \vcenter{\hbox{\includegraphics[scale=1]{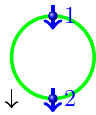}}}\ 
\right)
&=
c_{1\downarrow}^\dagger c_{2\downarrow}^\dagger
\Psi\left(
\ \vcenter{\hbox{\includegraphics[scale=1]{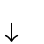}}}\ 
\right).
\end{align}

(3) $F$ move.
One of the most famous and nontrivial local moves is the $F$ move that may change the connecting topology of the domain walls. Here is the $F$ move with up spins at the bottom-left corner:
\begin{align}\label{F}
\Psi\left(
\ \vcenter{\hbox{\includegraphics[scale=1]{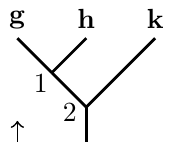}}}\ 
\right)
=
\nu_3(\mathbf{g,h,k})
(c_{1\uparrow}^\dagger)^{n_2(\mathbf{g,h})}
(c_{2\uparrow}^\dagger)^{n_2(\mathbf{gh,k})}
(c_{4\uparrow})^{n_2(\mathbf{g,hk})}
(c_{3\s_3})^{n_2(\mathbf{h,k})}
\Psi\left(
\ \vcenter{\hbox{\includegraphics[scale=1]{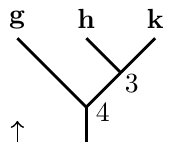}}}\ 
\right).
\end{align}
For the four fermions $c_{i\s_i}$ ($i=1,2,3,4$), the spin $\s_i$ are determined by the decoration rules given in the main text. In particular, $\s_3=\uparrow$ ($\s_3=\downarrow$) if the domain wall $\mathbf{g}=e$ ($\mathbf{g}=T$) is trivial (nontrivial). And $n_2(\mathbf{g,h})=0,1$ indicates whether the fermion is decorated or not, depending on the domain wall configurations $\mathbf{g}$ and $\mathbf{h}$. The factor $\nu_3(\mathbf{g,h,k})$ is a complex number that plays the same rule as the $F$ symbol in the bosonic system.

Under the time reversal symmetry action, the $F$ move \eq{F} becomes
\begin{align}\label{Fdown}
\Psi\left(
\ \vcenter{\hbox{\includegraphics[scale=1]{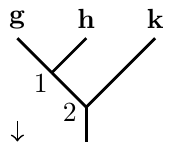}}}\ 
\right)
=
\nu_3^\ast(\mathbf{g,h,k})
(-1)^{\delta_{\mathbf{g},T}n_2(\mathbf{h,k})}
(c_{1\downarrow}^\dagger)^{n_2(\mathbf{g,h})}
(c_{2\downarrow}^\dagger)^{n_2(\mathbf{gh,k})}
(c_{4\downarrow})^{n_2(\mathbf{g,hk})}
(c_{3,-\s_3})^{n_2(\mathbf{h,k})}
\Psi\left(
\ \vcenter{\hbox{\includegraphics[scale=1]{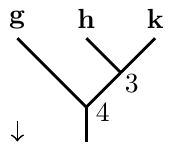}}}\ 
\right),
\end{align}
where all the bosonic and fermionic spins are flipped. In particular, the spin at the bottom-left corner is now pointing down. If and only if $\s_3=\downarrow$ and $n_2(\mathbf{h,k})=1$, i.e., the position $3$ is decorated by a spin-down fermion, then the symmetry action will result in an additional minus sign from $c_{3\downarrow}\rightarrow -c_{3\uparrow}$. This is the origin of the sign $(-1)^{\delta_{\mathbf{g},T}n_2(\mathbf{h,k})}$ after the complex conjugation of $\nu_3$. This is exactly the symmetry phase factor $e^{2\pi i \om_2\smile n_2}$ in \eq{symm_sign} of the main text.

As the most nontrivial special case of the $F$ move, the one with $\mathbf{g=h=k}=T$ will reconnect the domain walls. Depending on the spin at the bottom-left corner, there are two of them as symmetry partner:
\begin{align}\label{F_1}
\Psi\left(
\ \vcenter{\hbox{\includegraphics[scale=1]{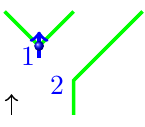}}}\ 
\right)
&=
c_{1\uparrow}^\dagger c_{3\downarrow}
\Psi\left(
\ \vcenter{\hbox{\includegraphics[scale=1]{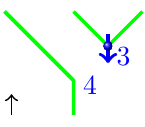}}}\ 
\right),\\\label{F_2}
\Psi\left(
\ \vcenter{\hbox{\includegraphics[scale=1]{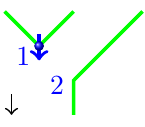}}}\ 
\right)
&=
-c_{1\downarrow}^\dagger c_{3\uparrow}
\Psi\left(
\ \vcenter{\hbox{\includegraphics[scale=1]{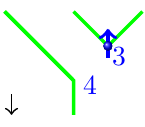}}}\ 
\right).
\end{align}
Since there is only one down spin, there is a minus sign difference between these two equations. It is exactly the sign $(-1)^{\delta_{\mathbf{g},T}n_2(\mathbf{h,k})}$ in \eq{Fdown}. The above two $F$ moves can be summarized as
\begin{align}
\Psi\left(
\ \vcenter{\hbox{\includegraphics[scale=1]{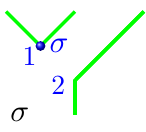}}}\ 
\right)
&=
\s c_{1\s}^\dagger c_{3\bar\s}
\Psi\left(
\ \vcenter{\hbox{\includegraphics[scale=1]{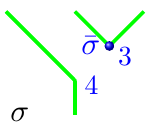}}}\ 
\right),
\end{align}
where the bottom-left corner spin is $\s=\uparrow/\downarrow=+/-$, and $\bar\s=-\s$ is the inverse of $\s$.

(4) Domain wall bending.
Using the above several basic moves, we can derive another useful local move that create or annihilate a pair of minimum and maximum points of a domain wall. For instance, we can show that
\begin{align}\label{bending}
\Psi\left(
\ \vcenter{\hbox{\includegraphics[scale=1]{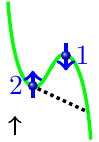}}}\ 
\right)
=
c_{2\uparrow}^\dagger c_{3\downarrow}
\Psi\left(
\ \vcenter{\hbox{\includegraphics[scale=1]{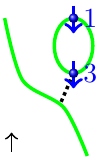}}}\ 
\right)
=
c_{2\uparrow}^\dagger c_{3\downarrow}
c_{1\downarrow}^\dagger c_{3\downarrow}^\dagger
\Psi\left(
\ \vcenter{\hbox{\includegraphics[scale=1]{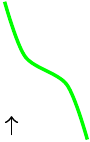}}}\ 
\right)
=
c_{1\downarrow}^\dagger c_{2\uparrow}^\dagger
\Psi\left(
\ \vcenter{\hbox{\includegraphics[scale=1]{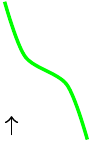}}}\ 
\right).
\end{align}
The first and second steps come from the $F$ move \eq{F} and the move \eq{Odown}. Under the time reversal symmetry action, the above equation becomes
\begin{align}\label{bending2}
\Psi\left(
\ \vcenter{\hbox{\includegraphics[scale=1]{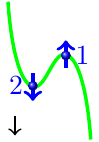}}}\ 
\right)
=
-c_{1\uparrow}^\dagger c_{2\downarrow}^\dagger
\Psi\left(
\ \vcenter{\hbox{\includegraphics[scale=1]{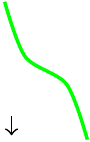}}}\ 
\right),
\end{align}
with an additional minus sign from the symmetry action on the spin-down fermion. There is another bending direction. We can similarly derive these two moves as
\begin{align}
\Psi\left(
\ \vcenter{\hbox{\includegraphics[scale=1]{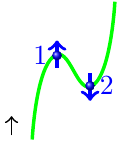}}}\ 
\right)
&=
c_{1\uparrow}^\dagger c_{2\downarrow}^\dagger
\Psi\left(
\ \vcenter{\hbox{\includegraphics[scale=1]{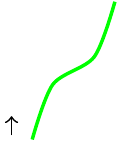}}}\ 
\right),\\
\Psi\left(
\ \vcenter{\hbox{\includegraphics[scale=1]{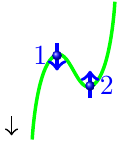}}}\ 
\right)
&=
-c_{1\uparrow}^\dagger c_{2\downarrow}^\dagger
\Psi\left(
\ \vcenter{\hbox{\includegraphics[scale=1]{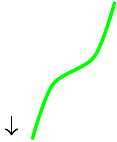}}}\ 
\right).
\end{align}


\subsection{B.2. General wavefunction}

The ground-state wavefunction \eq{TIwf} is a superposition of all possible bosonic spin configurations with fermionic $\Uf$ charge decorations. Using the basic local moves discussed above, we can derive the coefficient $\Psi(c)$ for any given configuration $c$, from the no-domain-wall configuration $c_0$. Since the basic moves satisfy the consistency conditions such as super-cocycle equation, the final result is independent of the paths we choose from the configuration $c$ to $c_0$.

As an example, let us consider the following configuration with $N$ decorated fermions ($N$ is always even). We can use \eq{bending} iteratively to cancel pairs of minimum and maximum points of the domain wall.
\begin{equation}
\begin{split}
\Psi\left(
\ \vcenter{\hbox{\includegraphics[scale=1]{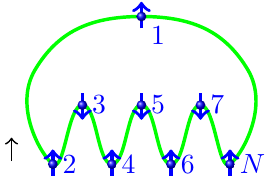}}}\ 
\right)
&=
c_{3\downarrow}^\dagger c_{2\uparrow}^\dagger
\Psi\left(
\ \vcenter{\hbox{\includegraphics[scale=1]{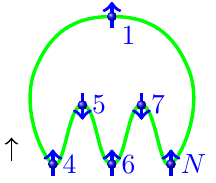}}}\ 
\right)
=
(c_{3\downarrow}^\dagger c_{2\uparrow}^\dagger)
(c_{5\downarrow}^\dagger c_{4\uparrow}^\dagger)
\Psi\left(
\ \vcenter{\hbox{\includegraphics[scale=1]{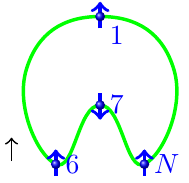}}}\ 
\right)\\
&=...
=
(c_{3\downarrow}^\dagger c_{2\uparrow}^\dagger)
(c_{5\downarrow}^\dagger c_{4\uparrow}^\dagger)
...
(c_{N-1\downarrow}^\dagger c_{N-2\uparrow}^\dagger)
\Psi\left(
\ \vcenter{\hbox{\includegraphics[scale=1]{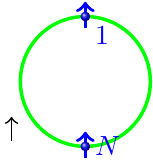}}}\ 
\right)
\\
&=
(c_{3\downarrow}^\dagger c_{2\uparrow}^\dagger)
(c_{5\downarrow}^\dagger c_{4\uparrow}^\dagger)
...
(c_{N-1\downarrow}^\dagger c_{N-2\uparrow}^\dagger)
(c_{1\uparrow}^\dagger c_{N\uparrow}^\dagger)
\Psi\left(
\ \vcenter{\hbox{\includegraphics[scale=1]{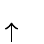}}}\ 
\right)
\\
&=
(-1)^{N/2-1}
c_{1\uparrow}^\dagger c_{2\uparrow}^\dagger c_{3\downarrow}^\dagger ... c_{N\uparrow}^\dagger
\Psi\left(
\ \vcenter{\hbox{\includegraphics[scale=1]{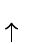}}}\ 
\right)
\end{split}
\end{equation}
And the last domain wall loop can be annihilated using \eq{O}. In this way, we relate the coefficient of the original configuration to that of the no-domain-wall configuration (which can be simplify set to $1$ before normalization). It seems that there is no compact expression for the coefficient of arbitrary domain wall configuration. But the coefficient is only $\pm 1$ apart from a sequence of fermion creation operators.

\section{B. Commuting-projector Hamiltonian for 2D interacting TI}
\label{App:H}

Similar to any fixed-point wavefunction of topological phases, we can construct an exactly-solvable lattice Hamiltonian for the ground-state wavefunction \eq{TIwf} of the 2D interacting topological insulator protected by $\Uf$ charge and time reversal symmetries. Here we will present the details of this Hamiltonian.

As shown in \eq{H} of the main text, the commuting-projector parent Hamiltonian reads
\begin{align}
H = - \sum_\bigtriangleup P_\bigtriangleup - \sum_\bigtriangledown P_\bigtriangledown - \sum_{\text{site }s} \frac{1+A_s}{2} \prod_\bigtriangleup P_\bigtriangleup \prod_\bigtriangledown P_\bigtriangledown.
\end{align}
The two triangle terms $P_\bigtriangleup$ and $P_\bigtriangledown$ enforce the fermionic $\Uf$ charge decoration rules for each triangle of the triangular lattice. The only four spin configurations with $\Uf$ charge decorations are
\begin{align}
\ \vcenter{\hbox{\includegraphics[scale=1]{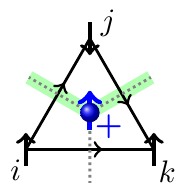}}}\ 
\ \vcenter{\hbox{\includegraphics[scale=1]{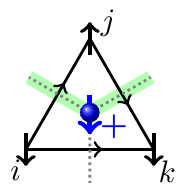}}}\ 
\ \vcenter{\hbox{\includegraphics[scale=1]{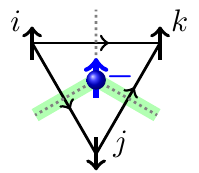}}}\ 
\ \vcenter{\hbox{\includegraphics[scale=1]{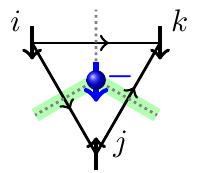}}}\ 
.
\end{align}
For convenience, we choose a branching structure (orientations of all links) of the triangle lattice shown in \fig{fig:H}. And the three vertices of each triangle $\langle ijk\rangle$ are chosen such that the link orientations are $i\rightarrow j$, $j\rightarrow k$ and $i\rightarrow k$.
In this notation, a spin-$\s_i$ fermion is decorated at the center of the triangle $\langle ijk\rangle$ if and only if $\s_i=-\s_j=\s_k$. The triangle projectors $P_{\bigtriangleup/\bigtriangledown}$ has the following expression
\begin{align}\nonumber
P_{\bigtriangleup/\bigtriangledown}
&=
\begin{cases}
n_{ijk}^{\s_i} (1-n_{ijk}^{-\s_i}), & \s_i=-\s_j=\s_k\\
(1-n_{ijk}^{\s_i}) (1-n_{ijk}^{-\s_i}), &\mathrm{others}
\end{cases}
\\
&=
n_{ijk}^{\s_i} (1-n_{ijk}^{-\s_i}) \left|\frac{\s_i-\s_j}{2}\right| \left|\frac{\s_j-\s_k}{2}\right|
+(1-n_{ijk}^{\uparrow}) (1-n_{ijk}^{\downarrow}) \left(1-\left|\frac{\s_i-\s_j}{2}\right| \left|\frac{\s_j-\s_k}{2}\right|\right),
\end{align}
where $n_{ijk}^\s := (c_{ijk}^\s)^\dagger c_{ijk}^\s$ is the fermion number operator for the fermion of the triangle $\langle ijk\rangle$ with spin $\s$.


\begin{table}[th!]
\begin{tabular}{|c|c|c||c|c|c|}
\hline
configurations
& $\s_1,...,\s_6$
& $A_s$
& configurations
& $\s_1,...,\s_6$
& $A_s$\\
\hline\hline
$
\vcenter{\hbox{\includegraphics[scale=1]{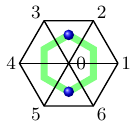}}}
\!\!\leftarrow\!\!
\vcenter{\hbox{\includegraphics[scale=1]{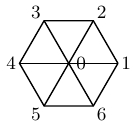}}}
$
& $+$$+$$+$$+$$+$$+$
& $c_{302}^{\s_3\dagger} c_{506}^{\s_5\dagger} X_0$
&
$
\vcenter{\hbox{\includegraphics[scale=1]{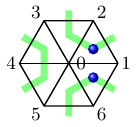}}}
\!\!\leftarrow\!\!
\vcenter{\hbox{\includegraphics[scale=1]{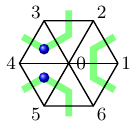}}}
$
& $-$$+$$-$$+$$-$$+$
& $c_{021}^{\s_0\dagger} c_{061}^{\s_0\dagger} X_0 c_{430}^{\s_4} c_{450}^{\s_4}$
\\
\hline
\hline
$
\vcenter{\hbox{\includegraphics[scale=1]{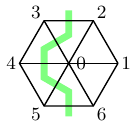}}}
\!\!\leftarrow\!\!
\vcenter{\hbox{\includegraphics[scale=1]{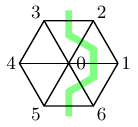}}}
$
& $-$$-$$+$$+$$+$$-$
& $X_0$
&
$
\vcenter{\hbox{\includegraphics[scale=1]{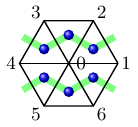}}}
\!\!\leftarrow\!\!
\vcenter{\hbox{\includegraphics[scale=1]{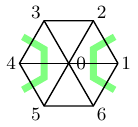}}}
$
& $-$$+$$+$$-$$+$$+$
& $Z_4 c_{302}^{\s_3\dagger} c_{430}^{\s_4\dagger} c_{450}^{\s_4\dagger} c_{506}^{\s_5\dagger} c_{021}^{\s_0\dagger} c_{061}^{\s_0\dagger} X_0$
\\
\hline
\hline
$
\vcenter{\hbox{\includegraphics[scale=1]{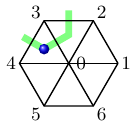}}}
\!\!\leftarrow\!\!
\vcenter{\hbox{\includegraphics[scale=1]{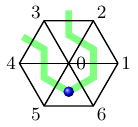}}}
$
& $-$$-$$+$$-$$-$$-$
& $c_{430}^{\s_4\dagger}  X_0 c_{506}^{\s_5}$
&
$
\vcenter{\hbox{\includegraphics[scale=1]{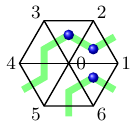}}}
\!\!\leftarrow\!\!
\vcenter{\hbox{\includegraphics[scale=1]{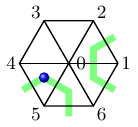}}}
$
& $-$$+$$+$$+$$-$$+$
& $c_{302}^{\s_3\dagger} c_{021}^{\s_0\dagger} c_{061}^{\s_0\dagger} X_0 c_{450}^{\s_4}$
\\
\hline
$
\vcenter{\hbox{\includegraphics[scale=1]{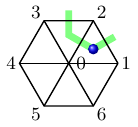}}}
\!\!\leftarrow\!\!
\vcenter{\hbox{\includegraphics[scale=1]{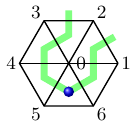}}}
$
& $-$$+$$-$$-$$-$$-$
& $c_{012}^{\s_0\dagger}  X_0 c_{506}^{\s_5}$
&
$
\vcenter{\hbox{\includegraphics[scale=1]{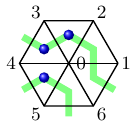}}}
\!\!\leftarrow\!\!
\vcenter{\hbox{\includegraphics[scale=1]{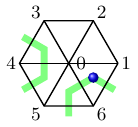}}}
$
& $+$$+$$+$$-$$+$$-$
& $c_{302}^{\s_3\dagger} c_{430}^{\s_4\dagger} c_{450}^{\s_4\dagger} X_0 c_{061}^{\s_0}$
\\
\hline
$
\vcenter{\hbox{\includegraphics[scale=1]{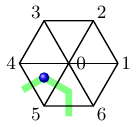}}}
\!\!\leftarrow\!\!
\vcenter{\hbox{\includegraphics[scale=1]{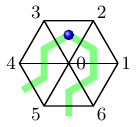}}}
$
& $-$$-$$-$$-$$+$$-$
& $c_{450}^{\s_4\dagger}  X_0 c_{302}^{\s_3}$
&
$
\vcenter{\hbox{\includegraphics[scale=1]{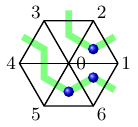}}}
\!\!\leftarrow\!\!
\vcenter{\hbox{\includegraphics[scale=1]{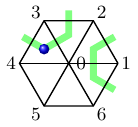}}}
$
& $-$$+$$-$$+$$+$$+$
& $c_{061}^{\s_0\dagger} c_{506}^{\s_5\dagger} c_{021}^{\s_0\dagger} X_0 c_{430}^{\s_4}$
\\
\hline
$
\vcenter{\hbox{\includegraphics[scale=1]{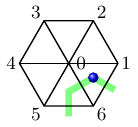}}}
\!\!\leftarrow\!\!
\vcenter{\hbox{\includegraphics[scale=1]{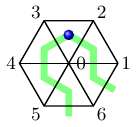}}}
$
& $-$$-$$-$$-$$-$$+$
& $c_{061}^{\s_0\dagger}  X_0 c_{302}^{\s_3}$
&
$
\vcenter{\hbox{\includegraphics[scale=1]{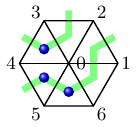}}}
\!\!\leftarrow\!\!
\vcenter{\hbox{\includegraphics[scale=1]{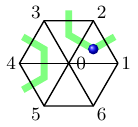}}}
$
& $+$$-$$+$$-$$+$$+$
& $c_{450}^{\s_4\dagger} c_{506}^{\s_5\dagger} c_{430}^{\s_4\dagger} X_0 c_{021}^{\s_0}$
\\
\hline
\hline
$
\vcenter{\hbox{\includegraphics[scale=1]{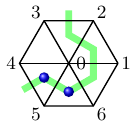}}}
\!\!\leftarrow\!\!
\vcenter{\hbox{\includegraphics[scale=1]{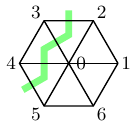}}}
$
& $+$$+$$-$$-$$+$$+$
& $Z_4 c_{450}^{\s_4\dagger} c_{506}^{\s_5\dagger} X_0$
&
$
\vcenter{\hbox{\includegraphics[scale=1]{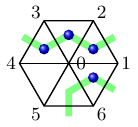}}}
\!\!\leftarrow\!\!
\vcenter{\hbox{\includegraphics[scale=1]{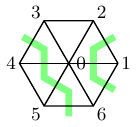}}}
$
& $-$$+$$+$$-$$-$$+$
& $c_{302}^{\s_3\dagger} c_{430}^{\s_4\dagger} c_{021}^{\s_0\dagger} c_{061}^{\s_0\dagger} X_0$
\\
\hline
$
\vcenter{\hbox{\includegraphics[scale=1]{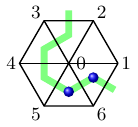}}}
\!\!\leftarrow\!\!
\vcenter{\hbox{\includegraphics[scale=1]{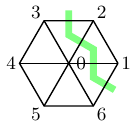}}}
$
& $-$$-$$+$$+$$+$$+$
& $Z_4 c_{061}^{\s_0\dagger} c_{506}^{\s_5\dagger} X_0$
&
$
\vcenter{\hbox{\includegraphics[scale=1]{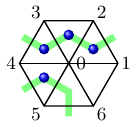}}}
\!\!\leftarrow\!\!
\vcenter{\hbox{\includegraphics[scale=1]{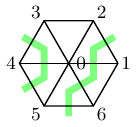}}}
$
& $-$$+$$+$$-$$+$$-$
& $c_{302}^{\s_3\dagger} c_{430}^{\s_4\dagger} c_{021}^{\s_0\dagger} c_{450}^{\s_4\dagger} X_0$
\\
\hline
$
\vcenter{\hbox{\includegraphics[scale=1]{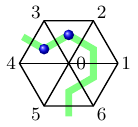}}}
\!\!\leftarrow\!\!
\vcenter{\hbox{\includegraphics[scale=1]{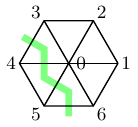}}}
$
& $+$$+$$+$$-$$-$$+$
& $Z_4 c_{302}^{\s_3\dagger} c_{430}^{\s_4\dagger} X_0$
&
$
\vcenter{\hbox{\includegraphics[scale=1]{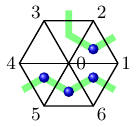}}}
\!\!\leftarrow\!\!
\vcenter{\hbox{\includegraphics[scale=1]{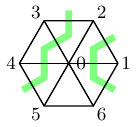}}}
$
& $-$$+$$-$$-$$+$$+$
& $c_{450}^{\s_4\dagger} c_{506}^{\s_5\dagger} c_{021}^{\s_0\dagger} c_{061}^{\s_0\dagger} X_0$
\\
\hline
$
\vcenter{\hbox{\includegraphics[scale=1]{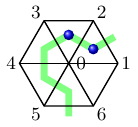}}}
\!\!\leftarrow\!\!
\vcenter{\hbox{\includegraphics[scale=1]{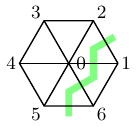}}}
$
& $-$$+$$+$$+$$+$$-$
& $Z_4 c_{302}^{\s_3\dagger} c_{021}^{\s_0\dagger} X_0$
&
$
\vcenter{\hbox{\includegraphics[scale=1]{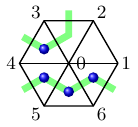}}}
\!\!\leftarrow\!\!
\vcenter{\hbox{\includegraphics[scale=1]{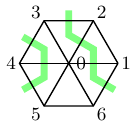}}}
$
& $-$$-$$+$$-$$+$$+$
& $c_{450}^{\s_4\dagger} c_{506}^{\s_5\dagger} c_{430}^{\s_4\dagger} c_{061}^{\s_0\dagger} X_0$
\\
\hline
\hline
$
\vcenter{\hbox{\includegraphics[scale=1]{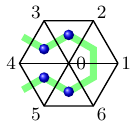}}}
\!\!\leftarrow\!\!
\vcenter{\hbox{\includegraphics[scale=1]{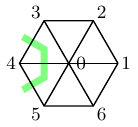}}}
$
& $+$$+$$+$$-$$+$$+$
& $c_{302}^{\s_3\dagger} c_{430}^{\s_4\dagger} c_{450}^{\s_4\dagger} c_{506}^{\s_5\dagger} X_0$
&
$
\vcenter{\hbox{\includegraphics[scale=1]{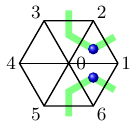}}}
\!\!\leftarrow\!\!
\vcenter{\hbox{\includegraphics[scale=1]{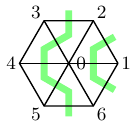}}}
$
& $-$$+$$-$$-$$-$$+$
& $Z_4 c_{021}^{\s_0\dagger} c_{061}^{\s_0\dagger}  X_0$
\\
\hline
$
\vcenter{\hbox{\includegraphics[scale=1]{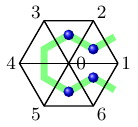}}}
\!\!\leftarrow\!\!
\vcenter{\hbox{\includegraphics[scale=1]{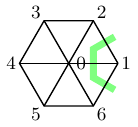}}}
$
& $-$$+$$+$$+$$+$$+$
& $c_{302}^{\s_3\dagger} c_{021}^{\s_0\dagger} c_{061}^{\s_0\dagger} c_{506}^{\s_5\dagger} X_0$
&
$
\vcenter{\hbox{\includegraphics[scale=1]{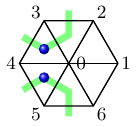}}}
\!\!\leftarrow\!\!
\vcenter{\hbox{\includegraphics[scale=1]{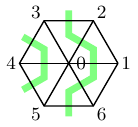}}}
$
& $-$$-$$+$$-$$+$$-$
& $Z_4 c_{430}^{\s_4\dagger} c_{450}^{\s_4\dagger}  X_0$
\\
\hline
\hline
$
\vcenter{\hbox{\includegraphics[scale=1]{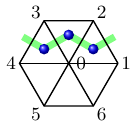}}}
\!\!\leftarrow\!\!
\vcenter{\hbox{\includegraphics[scale=1]{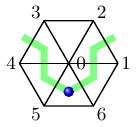}}}
$
& $-$$+$$+$$-$$-$$-$
& $Z_4 c_{302}^{\s_3\dagger} c_{430}^{\s_4\dagger} c_{021}^{\s_0\dagger} X_0 c_{506}^{\s_5}$
&
$
\vcenter{\hbox{\includegraphics[scale=1]{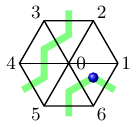}}}
\!\!\leftarrow\!\!
\vcenter{\hbox{\includegraphics[scale=1]{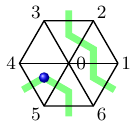}}}
$
& $-$$-$$+$$+$$-$$+$
& $Z_4 c_{061}^{\s_0\dagger} X_0 c_{450}^{\s_4}$
\\
\hline
$
\vcenter{\hbox{\includegraphics[scale=1]{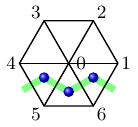}}}
\!\!\leftarrow\!\!
\vcenter{\hbox{\includegraphics[scale=1]{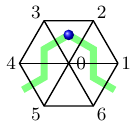}}}
$
& $-$$-$$-$$-$$+$$+$
& $Z_4 c_{450}^{\s_4\dagger} c_{506}^{\s_5\dagger} c_{061}^{\s_0\dagger} X_0 c_{302}^{\s_3}$
&
$
\vcenter{\hbox{\includegraphics[scale=1]{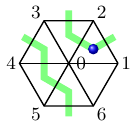}}}
\!\!\leftarrow\!\!
\vcenter{\hbox{\includegraphics[scale=1]{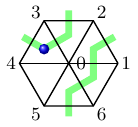}}}
$
& $-$$+$$-$$+$$+$$-$
& $Z_4 c_{021}^{\s_0\dagger} X_0 c_{430}^{\s_4}$
\\
\hline
\hline
$
\vcenter{\hbox{\includegraphics[scale=1]{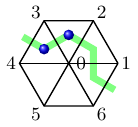}}}
\!\!\leftarrow\!\!
\vcenter{\hbox{\includegraphics[scale=1]{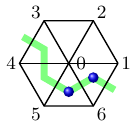}}}
$
& $+$$+$$+$$-$$-$$-$
& $c_{302}^{\s_3\dagger} c_{430}^{\s_4\dagger} X_0 c_{506}^{\s_5} c_{061}^{\s_0}$
&
$
\vcenter{\hbox{\includegraphics[scale=1]{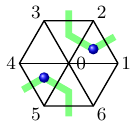}}}
\!\!\leftarrow\!\!
\vcenter{\hbox{\includegraphics[scale=1]{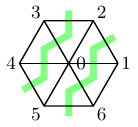}}}
$
& $-$$+$$-$$-$$+$$-$
& $Z_4 c_{021}^{\s_0\dagger} c_{450}^{\s_4\dagger}  X_0$
\\
\hline
$
\vcenter{\hbox{\includegraphics[scale=1]{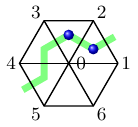}}}
\!\!\leftarrow\!\!
\vcenter{\hbox{\includegraphics[scale=1]{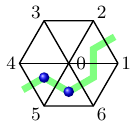}}}
$
& $-$$+$$+$$+$$-$$-$
& $c_{302}^{\s_3\dagger} c_{021}^{\s_0\dagger} X_0 c_{506}^{\s_5} c_{450}^{\s_4}$
&
$
\vcenter{\hbox{\includegraphics[scale=1]{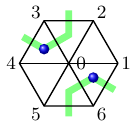}}}
\!\!\leftarrow\!\!
\vcenter{\hbox{\includegraphics[scale=1]{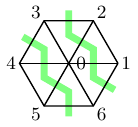}}}
$
& $-$$-$$+$$-$$-$$+$
& $Z_4 c_{430}^{\s_4\dagger} c_{061}^{\s_0\dagger}  X_0$
\\
\hline
\end{tabular}
\caption{Summary of the spin flipping term $A_s$. 
}
\label{tab:As}
\end{table}

The last term $(1+A_s)/2$ of the Hamiltonian is a projector that only acts nontrivially within the subspace with $P_\bigtriangleup = P_\bigtriangledown = 1$ for all triangles (see \tab{tab:As} for explicit expressions for $A_s$). In this subspace, the operator $A_s$ flips the spin at site $s$ and changes the fermionic $\Uf$ charge decorations accordingly. Since the six bosonic spins nearing site $s$ affect the domain wall configurations and charge decorations, the operator $A_s$ also depends on these spins. For a given domain wall configuration (and the corresponding legitimate charge decoration), we can write down the explicit expression of $A_s$. For example, when acting on the domain wall configuration without any fermionic charge decoration, $A_s$ is simply the Pauli operator $X_0$ of the spin at site $s$ which is numbered $0$ in the figure:
\begin{gather}
\ \vcenter{\hbox{\includegraphics[scale=1]{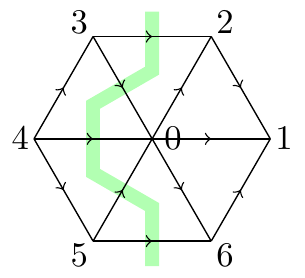}}}\ 
\xleftarrow{\ A_s\ }
\ \vcenter{\hbox{\includegraphics[scale=1]{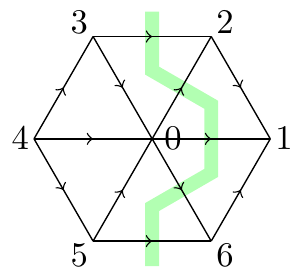}}}\ 
,\\
A_s = X_0.
\end{gather}
It corresponds to the local move of changing domain wall shape in \eq{shape}. So there is no fermion creation or annihilation procedure.
When we consider a configuration without any domain wall near the site $s$, the operator $A_s$ will create one with two fermionic $\Uf$ charges just as \eq{O}:
\begin{gather}
\ \vcenter{\hbox{\includegraphics[scale=1]{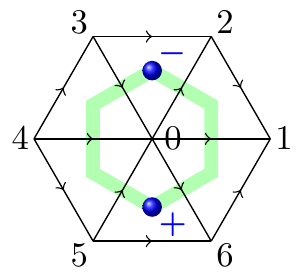}}}\ 
\xleftarrow{\ A_s\ }
\ \vcenter{\hbox{\includegraphics[scale=1]{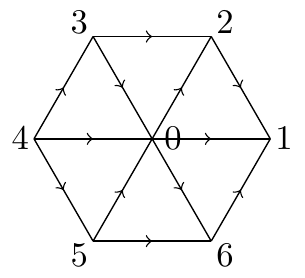}}}\ 
,\\
A_s = (c_{302}^{\s_3})^\dagger (c_{506}^{\s_5})^\dagger X_0.
\end{gather}
Similarly, the $F$ moves Eqs.~(\ref{F_1}) and (\ref{F_2}) can be put on the triangular lattice as
\begin{gather}
\ \vcenter{\hbox{\includegraphics[scale=1]{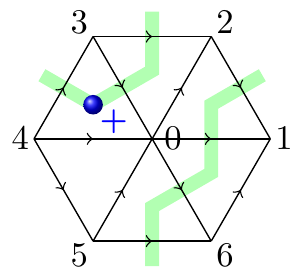}}}\ 
\xleftarrow{\ A_s\ }
\ \vcenter{\hbox{\includegraphics[scale=1]{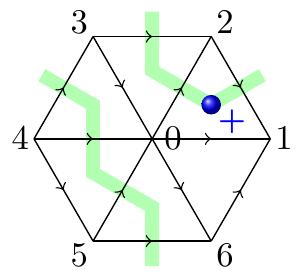}}}\ 
,\\
A_s = Z_5 (c_{430}^{\s_4})^\dagger X_0 c_{021}^{\s_0},
\end{gather}
where $Z_5$ is the Pauli-$Z$ operator acting on the spin at the bottom-left corner site $5$. The domain wall bending moves Eqs.~(\ref{bending}) and (\ref{bending2}) correspond to the following lattice 
\begin{gather}
\ \vcenter{\hbox{\includegraphics[scale=1]{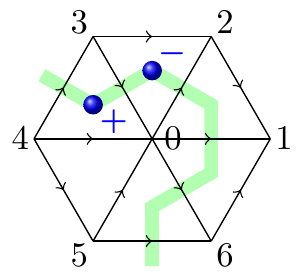}}}\ 
\xleftarrow{\ A_s\ }
\ \vcenter{\hbox{\includegraphics[scale=1]{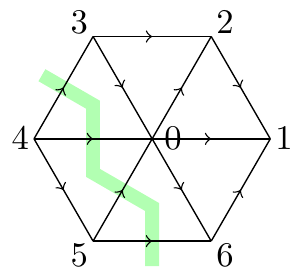}}}\ 
,\\
A_s = Z_5 (c_{302}^{\s_3})^\dagger (c_{430}^{\s_4})^\dagger X_0.
\end{gather}

\begin{gather}
\ \vcenter{\hbox{\includegraphics[scale=1]{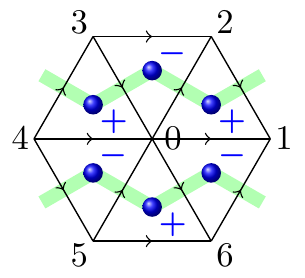}}}\ 
\xleftarrow{\ A_s\ }
\ \vcenter{\hbox{\includegraphics[scale=1]{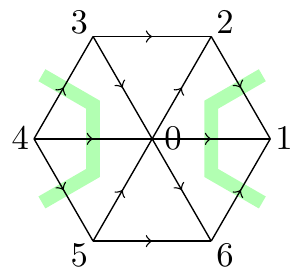}}}\ 
,\\
A_s = - Z_4
(c_{430}^{\s_4})^\dagger (c_{302}^{\s_3})^\dagger (c_{021}^{\s_0})^\dagger
(c_{450}^{\s_4})^\dagger (c_{506}^{\s_5})^\dagger (c_{061}^{\s_0})^\dagger
X_0.
\end{gather}

The full expressions for $A_s$ acting on all different spin configurations $\{\sigma_1,\sigma_2,...,\sigma_6\}$ surrounding the vertex $s$ are summarized in \tab{tab:As}. In this table, we assume that $A_s$ is acting on the right spin configuration $\{\s_0=+,\s_1,...,\s_6\}$, resulting in the left final spin configuration $\{\s_0=-,\s_1,...,\s_6\}$. The Hermitian conjugate of these operators will transform $s_0=-$ to $s_0=+$ with $\s_i$ ($1\le i \le 6$) fixed. In total, there are $2^6=64$ spin configurations $\{\sigma_i\}$ ($1\le i \le 6$). They correspond to $32$ domain wall configuration pairs listed in \tab{tab:As}.

\section{C. Nontrivial symmetry-breaking edge state of 2D interacting TI}

In the main text of the paper, we claim that the constructed FSPT with symmetry $G_f=\Uf\rtimes_{\rho_1,\om_2} \Z_2^T=(\Uf\rtimes \Z_4^T)/\Z_2$ is the interacting analog of the 2D time-reversal-invariant TI in the free-fermion system. This can be shown if we can construct the same helical gapless edge states as the free-fermion TI~\cite{Metlitski2019}. On the other hand, the 2D TI is also known to have nontrivial gapped, symmetry-breaking edge states if we break the $\U$ or $\Z_2^T$ symmetry by proximity effect on its boundary~\cite{Son2019}. In this section, we will show that the edge time-reversal-symmetry domain wall of the constructed model will trap a half $\Uf$ charge. Therefore, the interacting state we constructed indeed shares the same nontrivial edge property as the free-fermion TI.

\begin{figure}[h]
\centering
$\vcenter{\hbox{\includegraphics[scale=1]{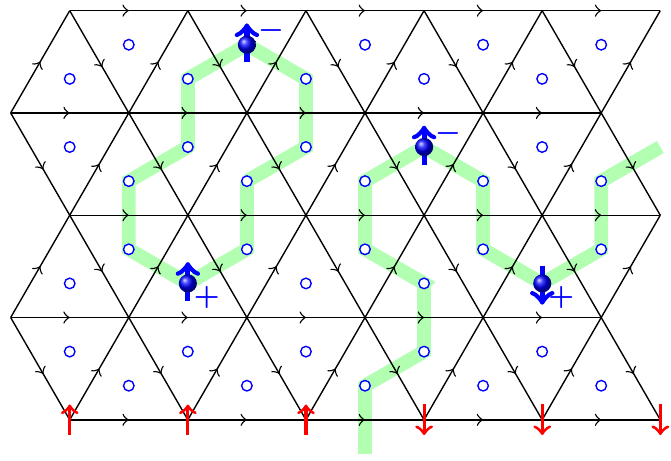}}}$
\caption{Boundary spin domain wall. We add a position-dependent boundary Zeeman field to polarize the boundary spins (red arrows), such that there is an edge domain wall.}
\label{fig:edge}
\end{figure}

Let us add a Zeeman field to the boundary of the constructed model, such that the left and right edge spins are pointing up and down, respectively (see \fig{fig:edge}). Although the bulk spins are fluctuating, the red spins on the boundary are fixed. And there is an spin domain wall crossing the boundary. Just as in the bosonic model~\cite{wang2021exactly}, the $\Uf$ charge inside a fixed triangle is nonzero in the vacuum state. So what we need to calculate is the relative charge of the domain-wall configuration and the non-domain-wall configuration on the boundary. For the edge state without domain wall, the wavefunction of an edge triangle looks like
\begin{align}
|\Psi\rangle_{\uparrow\uparrow} \sim
\Bigg|
\vcenter{\hbox{\includegraphics[scale=1]{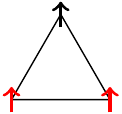}}}
\Bigg\rangle
+
\Bigg|
\vcenter{\hbox{\includegraphics[scale=1]{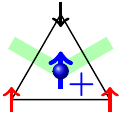}}}
\Bigg\rangle,
\end{align}
where the two red edge spins are fixed to up direction. The average $\Uf$ charge inside this triangle is then
\begin{align}
\langle Q \rangle_{\uparrow\uparrow} = 1/2.
\end{align}
On the other hand, if there is an edge domain wall, the wavefunction near the edge triangle is basically
\begin{align}
|\Psi\rangle_{\uparrow\downarrow} \sim
\Bigg|
\vcenter{\hbox{\includegraphics[scale=1]{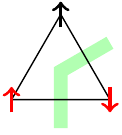}}}
\Bigg\rangle
+
\Bigg|
\vcenter{\hbox{\includegraphics[scale=1]{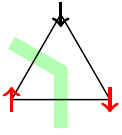}}}
\Bigg\rangle
\end{align}
And the average $\U$ charge is
\begin{align}
\langle Q \rangle_{\uparrow\downarrow} = 0.
\end{align}
Therefore, the relative charge of the edge domain wall is
\begin{align}
\Delta Q = \langle Q \rangle_{\uparrow\downarrow} - \langle Q \rangle_{\uparrow\uparrow} = -1/2.
\end{align}
This is exactly the half charge we expected on the boundary domain wall of the 2D TI with time reversal symmetry.


\section{D. From projective representation of $G$ to linear representation of $G_f$}

In this section, we will show that the symmetry transformation rules in \eq{Ugc} of the main text, i.e.,
\begin{align}\label{symm1}
U_\theta |g_i\rangle &= |g_i\rangle,\\\label{symm2}
U(g) |g_i\rangle &= |gg_i\rangle,\\\label{symm3}
U_\theta c_{ijk}^{\s,q} U_\theta^\dagger &= e^{- i q\theta} c_{ijk}^{\s,q},\\\label{symm4}
U(g) c_{ijk}^{\s,q} U(g)^\dagger &= e^{-2\pi i \om_2(g,\s) (-1)^{\rho_1(g)+s_1(g)}q} c_{ijk}^{g\s,(-1)^{\rho_1(g)+s_1(g)}q},
\end{align}
are linear representations of $G_f$ on these bosonic/fermionic degrees of freedom. The statement is true for arbitrary $G_f$, namely arbitrary $\rho_1,s_1\in H^1(G,\Z_2)$ and $\om_2\in H_{\rho_1}^2(G,\Uf)$ satisfying the cocycle conditions
\begin{align}
(\dd \rho_1)(g,h) &= \rho_1(h)-\rho_1(gh)+\rho_1(g) = 0\quad (\mathrm{mod}\ 2),\\
(\dd s_1)(g,h) &= s_1(h)-s_1(gh)+s(g) = 0\quad (\mathrm{mod}\ 2),\\\label{dom2}
(\dd_{\rho_1}\om_2)(g,h,k) &= (-1)^{\rho_1(g)}\om_2(h,k) - \om_2(gh,k) + \om_2(g,hk) - \om_2(g,h) = 0\quad (\mathrm{mod}\ 1).
\end{align}

From \eq{symm1}, the bosonic spins with basis $|g_i\rangle$ do not carry any nontrivial charge of $\Uf$. So the linear transformation of $G$ in \eq{symm2} also gives a $|G|$-dimensional linear representation of $G_f$ on the bosonic spins.

The boson/fermion $c_{ijk}^{\s,q}$ has $\Uf$ charge $q$ from \eq{symm3}. Combing it with the complicated transformation rule \eq{symm4} under $G$, we can obtain the transformation of $c_{ijk}^{\s,q}$ under arbitrary $e^{2\pi i \theta Q}g\in G_f$ as
\begin{align}\nonumber
U(e^{2\pi i \theta Q}g) c_{ijk}^{\s,q} U(e^{2\pi i \theta Q}g)^\dagger
&=
e^{2\pi i \theta Q}U(g) c_{ijk}^{\s,q} U(g)^\dagger e^{-2\pi i \theta Q}\\\nonumber
&=
e^{2\pi i \theta Q}
e^{-2\pi i \om_2(g,\s) (-1)^{\rho_1(g)+s_1(g)}q} c_{ijk}^{g\s,(-1)^{\rho_1(g)+s_1(g)}q}
e^{-2\pi i \theta Q}\\
&=
e^{-2\pi i [\om_2(g,\s)+\theta] (-1)^{\rho_1(g)+s_1(g)}q}
c_{ijk}^{g\s,(-1)^{\rho_1(g)+s_1(g)}q}.
\end{align}
Now we apply the action of another element $e^{2\pi i \phi Q}h\in G_f$. The successive symmetry actions on $c_{ijk}^{\s,q}$ read
\begin{align}\nonumber\label{rep_action1}
&\quad\ 
U(e^{2\pi i \phi Q}h)
U(e^{2\pi i \theta Q}g) c_{ijk}^{\s,q} U(e^{2\pi i \theta Q}g)^\dagger
U(e^{2\pi i \phi Q}h)^\dagger\\\nonumber
&=
U(e^{2\pi i \phi Q}h)
e^{-2\pi i [\om_2(g,\s)+\theta] (-1)^{\rho_1(g)+s_1(g)}q}
c_{ijk}^{g\s,(-1)^{\rho_1(g)+s_1(g)}q}
U(e^{2\pi i \phi Q}h)^\dagger\\\nonumber
&=
e^{2\pi i \phi Q}
e^{-2\pi i [\om_2(g,\s)+\theta] (-1)^{\rho_1(g)+s_1(g)+s_1(h)}q}
U(h)
c_{ijk}^{g\s,(-1)^{\rho_1(g)+s_1(g)}q}
U(h)^\dagger
e^{-2\pi i \phi Q}\\\nonumber
&=
e^{2\pi i \phi Q}
e^{-2\pi i [\om_2(g,\s)+\theta] (-1)^{\rho_1(g)+s_1(hg)}q}
e^{-2\pi i \om_2(h,g\s) (-1)^{\rho_1(hg)+s_1(hg)}q}
c_{ijk}^{hg\s,(-1)^{\rho_1(hg)+s_1(hg)}q}
e^{-2\pi i \phi Q}\\\nonumber
&=
e^{-2\pi i [\om_2(g,\s)+\theta] (-1)^{\rho_1(g)+s_1(hg)}q}
e^{-2\pi i \om_2(h,g\s) (-1)^{\rho_1(hg)+s_1(hg)}q}
e^{2\pi i \phi Q}
c_{ijk}^{hg\s,(-1)^{\rho_1(hg)+s_1(hg)}q}
e^{-2\pi i \phi Q}\\
&=
e^{-2\pi i [\om_2(g,\s)+\theta] (-1)^{\rho_1(g)+s_1(hg)}q}
e^{-2\pi i \om_2(h,g\s) (-1)^{\rho_1(hg)+s_1(hg)}q}
e^{-2\pi i \phi (-1)^{\rho_1(hg)+s_1(hg)}q}
c_{ijk}^{hg\s,(-1)^{\rho_1(hg)+s_1(hg)}q},
\end{align}
where we used the cocycle conditions of $\rho_1$ and $s_1$.
On the other hand, if the $c_{ijk}^{\s,q}$'s support a linear representation of $G_f$, the successive actions of $e^{2\pi i \phi Q}h$ and $e^{2\pi i \theta Q}g$ should be the same as the action of
\begin{align}
e^{2\pi i \phi Q}h \times e^{2\pi i \theta Q}g
=
e^{2\pi i \phi Q + 2\pi i \theta (-1)^{\rho_1(h)} Q + 2\pi i \om_2(h,g) Q} hg.
\end{align}
We can calculate directly that
\begin{align}\nonumber\label{rep_action2}
&\quad\ 
U(e^{2\pi i \phi Q}h \times e^{2\pi i \theta Q}g)
c_{ijk}^{\s,q}
U(e^{2\pi i \phi Q}h \times e^{2\pi i \theta Q}g)^\dagger\\\nonumber
&=
U(e^{2\pi i \phi Q + 2\pi i \theta (-1)^{\rho_1(h)} Q + 2\pi i \om_2(h,g) Q} hg)
c_{ijk}^{\s,q}
U(e^{2\pi i \phi Q + 2\pi i \theta (-1)^{\rho_1(h)} Q + 2\pi i \om_2(h,g) Q} hg)^\dagger\\\nonumber
&=
e^{2\pi i \left[\phi+\theta (-1)^{\rho_1(h)}+\om_2(h,g)\right] Q}
U(hg)
c_{ijk}^{\s,q}
U(hg)^\dagger
e^{-2\pi i \left[\phi+\theta (-1)^{\rho_1(h)}+\om_2(h,g)\right] Q}\\\nonumber
&=
e^{2\pi i \left[\phi+\theta (-1)^{\rho_1(h)}+\om_2(h,g)\right] Q}
e^{-2\pi i \om_2(hg,\s) (-1)^{\rho_1(hg)+s_1(hg)}q} c_{ijk}^{hg\s,(-1)^{\rho_1(hg)+s_1(hg)}q}
e^{-2\pi i \left[\phi+\theta (-1)^{\rho_1(h)}+\om_2(h,g)\right] Q}\\
&=
e^{-2\pi i \left[\phi+\theta (-1)^{\rho_1(h)}+\om_2(h,g)\right] (-1)^{\rho_1(hg)+s_1(hg)}q}
e^{-2\pi i \om_2(hg,\s) (-1)^{\rho_1(hg)+s_1(hg)}q}
c_{ijk}^{hg\s,(-1)^{\rho_1(hg)+s_1(hg)}q}.
\end{align}
Using the cocycle condition $(\dd_{\rho_1}\om_2)(h,g,\s)=0$ from \eq{dom2}, one can easily check that the results of \eqs{rep_action1}{rep_action2} are exactly the same.
Therefore, we have
\begin{align}
U(e^{2\pi i \phi Q}h \times e^{2\pi i \theta Q}g)
=U(e^{2\pi i \phi Q}h) U(e^{2\pi i \theta Q}g)
\end{align}
when acting on the d.o.f. $c_{ijk}^{\s,q}$ for arbitrary $e^{2\pi i \phi Q}h, e^{2\pi i \theta Q}g \in G_f$.

In summary, the bosonic spin $|g_i\rangle$ supports a $|G|$-dimensional linear representation of $G_f$. Since the $G$-action only changes the spin $\s\in G$ and the sign of the charge $q$ of $c_{ijk}^{\s,q}$, the collection of $c_{ijk}^{\s,\pm q}$ for all $\s\in G$ supports a $2|G|$-dimensional linear representation of $G_f$. If $\rho_1=s_1$, the charge $q$ of $c_{ijk}^{\s,q}$ is unchanged under the $G$-action. So we have a $|G|$-dimensional linear representation of $G_f$ for each charge $q$. The examples of this simpler case include the interacting topological insulator protected by $G_f=(\Uf\rtimes \Z_4^T)/\Z_2$ discussed in details in the main text. This is the reason why we can fix the $+1$ ($-1$) charges for the up-pointing (down-pointing) triangles of the triangular lattice in the construction.

\section{E. Boundary anomalous SPT states to trivialize the bulk}

It is possible that the wavefunction we constructed from decoration can be connected to a trivial product state using fermionic symmetric local unitary transformations. If it happens, there is a 1D anomalous SPT states~\cite{WQG2018} on the boundary of this 2D wavefunction. The trivialization subgroups of the decoration data $(n_2,\nu_3)$ are given by the obstruction functions for 1D fermionic invertible states protected by $G_f=\Uf \rtimes_{\rho_1,\om_2} G$~\cite{WG20}.

The complex fermion decoration data $n_2$ could be trivialized by a 1D FSPT protected only by $\Uf$. However, there is no nontrivial 1D fermionic invertible states with $\Uf$ symmetry. Therefore, any nontrivial decoration $n_2\in H^2_{\rho_1+s_1}(G,\Z)$ will be a nontrivial state. This is different from the superconductor case where 1D Kitaev chain with anomalous $G_b$-action can be used to trivialize 2D $n_2$ decoration~\cite{WG20}.

There is another possible trivialization when we embed a BSPT into a fermionic system~\cite{GuWen2014}. The BSPT $\nu_3\in H^3_{s_1}(G,\U)$ may be trivialized by 1D ASPT state with fermionic $\Uf$ charge decorations. Similar to 2D, the decoration of this layer in 1D is specified by $n_1\in H^1_{\rho_1+s_1}(G,\Z)$. If $\rho_1=s_1$, the group $H^1_{\rho_1+s_1}(G,\Z)$ is always trivial as we assume $G$ to be finite. If $\rho_1\neq s_1$, on the other hand, the decoration group $H^1_{\rho_1+s_1}(G,\Z)$ is $\Z_2$. Using the same techniques to obtain the twisted super-cocycle equation in the main text, we can derive the consistency condition for 1D FSPT as
\begin{align}
\dd_{s_1}\nu_2 = e^{2\pi i \om_2\smile n_1},
\end{align}
where $\nu_2\in C_{s_1}^2(G,\U)/B_{s_1}^2(G,\U)$ is the wavefunction coefficient. So the trivialization subgroup for $\nu_3$ in 2D is
\begin{align}\label{gamma3}
\Gamma^3=\{e^{2\pi i \om_2\smile n_1}| n_1\in H^1_{\rho_1+s_1}(G,\Z)\} \subset H^3_{s_1}(G,\U),
\end{align}
which is trivial if $\rho_1=s_1$ and at most $\Z_2$ otherwise.

In summary, the 2D $G_f$-FSPT with decoration data $(n_2,\nu_3)$ [satisfying \eqs{dn2}{dnu3}] is a nontrivial FSPT, if and only if $n_2\in H^2_{\rho_1+s_1}(G,\Z)$ is a nontrivial cocycle, or $\nu_3\in C_{s_1}^3(G,\U)/B_{s_1}^3(G,\U)/\Gamma^3$ is nontrivial. The trivialization subgroup $\Gamma^3$ in \eq{gamma3} is associated with 1D anomalous SPT with $\Uf$ charge decorations.

\section{F. Relation to Gu-Wen supercohomology FSPT state}
The construction of fermionic insulators is closely related to the Gu-Wen group supercohomology SPT models, where similar decorations of complex fermions on domain walls are considered, but the $F$ move is only required to conserve the total fermion parity. Mathematically, the Gu-Wen supercohomology SPT construction relies again on two pieces of data, $\omega_2\in H^2(G, \Z_2)$ for the extension of the symmetry group by $G$, and $n_2\in H^2(G, \Z_2)$ for the complex fermion decoration. A natural question is which of the group supercohomology SPT phases can actually be realized in insulators. For simplicity, let us consider the case where $\omega_2$ is trivial, so the total symmetry group is $\Z_2^f\times G$. In this case, we have the following general result: the group supercohomology SPT phase labeled by $n_2$ is compactible with $\Uf$ symmetry if and only if $n_2$ is trivialized when it is canonically lifted to a cocycle in $H^2(G, \Uf)$. Physically, $n_2$ gives the projective reprenstation carried by a fermion parity flux. When the system has $\Uf$ symmetry, a fermion parity flux can be created by adiabatically inserting $\pi$ flux. However, if the projective representation on the $\pi$ flux requires a multi-dimensional representation space, the flux insertion can not be adiabatic, which is impossible. We present a more formal proof by showing that the gauged FSPT phase suffers from a 't Hooft anomaly between $\Uf$ and $G$ in this case. 

From the universal coefficient theorem, we have
\begin{equation}
	H^2(G, \Z_2)=\big(H^2(G, \Z)\otimes \Z_2\big)\oplus \mathrm{Tor}[H^3(G, \Z),\Z_2].
	\label{}
\end{equation}
The second term precisely gives those that can be canonically lifted to a $2$-cocycle in $H^2(G, \Uf)$. The first term corresponds to those $n_2$ which can be lifted to a $\Z$-coefficient cocycle, and therefore our construction applies.

One should be careful in comparing the classifications with or without $\Uf$, even though naively the former is a subgroup of the latter (which is actually true for unitary $G$ and trivial $\omega_2$). The reason is that the coboundaries are very different. For example, 2D TI becomes trivialized if $\Uf$ is broken down to $\Z_2^f$.

\section{G. Stacking group structure}

Since SPT states are short-range entangled, the stacking of two SPT states protected by the same symmetry would result in a new SPT state. There is also an inverse state for every SPT, such that the staking of them can be connected to a trivial product state. In this section, we will derive the stacking Abelian group structure of FSPT states constructed in the main text of the paper.

For every solution $(n_2,\nu_3)$ of the obstruction equations, we can construct a valid $G_f$-FSPT state using domain wall decorations. Now let us consider two states constructed from $(n_2,\nu_3)$ and $(n_2',\nu_3')$. Under stacking, the $\Uf$ charges $n_2$ and $n_2'\in H_{\rho_1+s_1}^2(G,\Z)$ would be combined in a triangle, resulting in a total charge
\begin{align}\label{N2}
N_2=n_2+n_2'.
\end{align}
This is the $\Uf$ charge decoration data for the stacked system.

The subtle part is the phase factor of the wavefunction. To obtain the coefficient of the stacked wavefunction, we have to consider the $F_\mathrm{tot}$ move of the combined system:
\begin{align}\label{2D:grp}
\vcenter{\hbox{\includegraphics[scale=1]{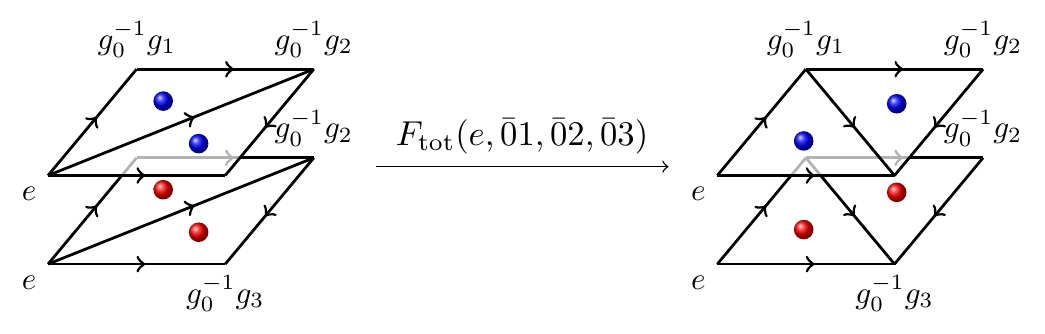}}}
.
\end{align}
We use blue and red balls to indicate the decorated $\Uf$ charges for the upper and lower layers. The  bosons/fermions in the two layers are denoted as $c_{ijk}^{\s,q}$ and $c_{ijk}'^{\s,q}$, respectively. Since the stacked system is the tensor product of the two layers, the total $F$ move is the tensor product of the $F$ moves for the two layers
\begin{align}\label{FxF}
F_\textrm{tot} = F\otimes F',
\end{align}
where the individual $F$ and $F'$ are defined as
\begin{align}\label{F_}
F(e,\bar 01,\bar 02,\bar 03) &= \nu_3(e,\bar 01,\bar 02,\bar 03) \big(c_{012}^{e,n_2(012)}\big)^\dagger \big(c_{023}^{e,n_2(023)}\big)^{\dagger} \big(c_{013}^{e,n_2(013)}\big) \big(c_{123}^{g_0^{-1}g_1,n_2(123)}\big),\\
F'(e,\bar 01,\bar 02,\bar 03) &= \nu_3'(e,\bar 01,\bar 02,\bar 03) \big({c'}_{012}^{e,n_2'(012)}\big)^\dagger \big({c'}_{023}^{e,n_2'(023)}\big)^{\dagger} \big({c'}_{013}^{e,n_2'(013)}\big) \big({c'}_{123}^{g_0^{-1}g_1,n_2'(123)}\big).
\end{align}

If we consider the combined system as a single layer, the $F_\mathrm{tot}$ symbol should be written as an operator acting on the four triangles sequentially. Using the triangle order convention similar to \eq{F_}, $F_\mathrm{tot}$ should have the form
\begin{align}\nonumber\label{Ftot}
F_\textrm{tot}(e,\bar 01,\bar 02,\bar 03)
&=
\mathcal V_3(e,\bar 01,\bar 02,\bar 03) 
\times
\left[\big(c_{012}^{e,n_2(012)}\big)^{\dagger}
\big(c_{012}'^{e,n_2'(012)}\big)^{\dagger}\right] 
\left[\big(c_{023}^{e,n_2(023)}\big)^{\dagger}
\big(c_{023}'^{e,n_2'(023)}\big)^{\dagger}\right]\\
&\quad\times
\left[c_{013}'^{e,n_2'(013)}
c_{013}^{e,n_2(013)} \right]
\left[c_{123}'^{g_0^{-1}g_1,n_2'(123)}
c_{123}^{g_0^{-1}g_1,n_2(123)}\right].
\end{align}
Here, we use the convention that the fermions in triangle $\langle ijk\rangle$ are created by $(c_{ijk}^{\s,q})^\dagger (c_{ijk}'^{\s})^\dagger$ for the two layers. Similarly, the annihilation operators $c_{ijk}'^{\s,q} c_{ijk}^{\s,q}$ are ordered as the complex conjugate.

To deform \eq{FxF} to \eq{Ftot}, we have to reorder the fermion creation and annihilation operators. Using the condition $\dd n_2=\dd n_2'=0$, it is easy to show that wavefunction of the combined system is
\begin{align}\label{V3}
\mathcal V_3 = \nu_3\nu_3'(-1)^{n_2\smile_1 n_2'},
\end{align}
where fermion signs appear as the cup-1 product defined by
\begin{align}
(n_2\smile_1 n_2')(0123) = n_2(023)n_2'(012) + n_2(013)n_2'(123),\quad (\text{mod }2).
\end{align}

We have obtained the stacking results of both the $\Uf$ charge \eq{N2} and the phase factor \eq{V3}. The new data $(N_2,\mathcal V_3)$ satisfies the new obstruction functions. In particular, one can show easily that
\begin{align}
\dd \mathcal V_3 = e^{2\pi i(\omega_2 \smile N_2 + \frac{1}{2}N_2\smile N_2) },
\end{align}
using the obstruction functions for $\nu_3$ and $\nu_3'$. Therefore, the stacking operation of $(n_2,\nu_3)$ and $(n_2',\nu_3')$ is
\begin{align}\label{grp}
(n_2,\nu_3) + (n_2',\nu_3') = (N_2,\mathcal V_3) := \left(n_2+n_2',\nu_3\nu_3'(-1)^{n_2\smile_1 n_2'}\right).
\end{align}
It can be also checked directly that the stacking operation satisfies all axioms of Abelian groups as expected.



\section{H. Examples of Wallpaper-Group Symmetries}

As examples of applying the results in the main text, we compute the classification of 2D interacting insulators protected by the wallpaper group symmetries.
Technically speaking, we are computing the classification of topological states protected by an onsite symmetry group, which has the same group structure as one of the 17 wallpaper groups.
However, according to the crystalline-equivalence principle, the classification results also applies to topological states protected by the actual wallpaper groups, with spatial actions.
Considering the fact that the symmetry group $G_f$ is an extension of the wallpaper group $G$ over $U(1)_f$,
it is important to notice that this correspondence between onsite and crystalline symmetry groups comes with a twist on the 2-cocycle $\omega_2$ characterizing this extension:
Let $\omega_2^{1/2}$ denote the 2-cocycle representing the extension for physical spin-$\frac12$ electrons transforming under wallpaper groups, onsite symmetries with $\omega_2=0$ ($\omega_2=\omega_2^{1/2}$) correspond to crystalline symmetries with $\omega_2=\omega_2^{1/2}$ ($\omega_2=0$), respectively.
In other words, spinless (spin-$\frac12$) electrons with onsite symmetries correspond to spin-$\frac12$ (spinless) electrons, respectively.
The results are computed using the algorithm in Ref.~\cite{resolution}.

\begin{table}
  \caption{Classification of 2D interacting-electron SPTs protected by 2D wallpaper groups, where fermions are spinless (spin-1/2 if the symmetry group is treated as spatial symmetries.)
  The answer is listed in terms of the $\Uf$ charge (C) and bosonic (B) layers.
  We notice that, when $\omega_2=0$, there is no trivialization $\Gamma^3$, so the column $B$ is given by $H^3(G, U(1)).$}
  \label{tab:2dwallpaper}
  \begin{tabularx}{\columnwidth}{CCCCC}
    \hline\hline
    SG & C & B \\
    \hline
    p1 & $\mathbb Z$ & 0 \\
    p2 & $3\mathbb Z_2\oplus\mathbb Z$ & $4\mathbb Z_2$\\
    p1m1 & $\mathbb Z_2\oplus\mathbb Z$ & $2\mathbb Z_2$\\
    p1g1 & $\mathbb Z$ & 0 \\
    c1m1 & $\mathbb Z$ & $\mathbb Z_2$ \\
    p2mm& $\mathbb Z$ & $8\mathbb Z_2$\\
    p2mg& $2\mathbb Z_2\oplus\mathbb Z$ & $3\mathbb Z_2$\\
    p2gg& $\mathbb Z_2\oplus\mathbb Z$ & $2\mathbb Z_2$\\
    c2mm& $\mathbb Z_2\oplus\mathbb Z$ & $5\mathbb Z_2$\\
    p4& $\mathbb Z_2\oplus\mathbb Z_4\oplus\mathbb Z$ & $\mathbb Z_2\oplus2\mathbb Z_4$ \\
    p4mm& $\mathbb Z_2\oplus\mathbb Z$ & $6\mathbb Z_2$\\
    p4gm& $\mathbb Z_2\oplus\mathbb Z$ & $2\mathbb Z_2\oplus\mathbb Z_4$\\
    p3& $2\mathbb Z_3\oplus\mathbb Z$ & $3\mathbb Z_3$\\
    p3m1& $2\mathbb Z_3\oplus\mathbb Z$ & $\mathbb Z_2$\\
    p31m& $\mathbb Z_3\oplus\mathbb Z$ & $\mathbb Z_6$\\
    p6& $\mathbb Z_6\oplus\mathbb Z$ & $2\mathbb Z_6$\\
    p6mm& $\mathbb Z_3\oplus\mathbb Z$& $4\mathbb Z_2$\\
    \hline\hline
  \end{tabularx}
\end{table}

\begin{table}
  \caption{Classification of 2D interacting-electron SPTs protected by 2D wallpaper groups, where fermions are spin-1/2 (spinless if the symmetry group is treated as spatial symmetries.)
  The answer is listed in terms of the $\Uf$ charge (C) and bosonic (B) layers.
  In the column labeled by B, we write $A\rightarrow B$, indicating that $A=H^3(G, U(1))$ is reduced to $B$ by trivialization $\Gamma^3$.}
  \label{tab:2dwallpaper}
  \begin{tabularx}{\columnwidth}{CCCCC}
    \hline\hline
    SG & C & B \\
    \hline
    p1 & $\mathbb Z$ & $0\rightarrow0$ \\
    p2 & $3\mathbb Z_2\oplus\mathbb Z$ & $4\mathbb Z_2\rightarrow4\mathbb Z_2$\\
    p1m1 & $\mathbb Z_2\oplus\mathbb Z$ & $2\mathbb Z_2\rightarrow2\mathbb Z_2$\\
    p1g1 & $\mathbb Z$ & $0\rightarrow0$ \\
    c1m1 & $\mathbb Z$ & $\mathbb Z_2\rightarrow\mathbb Z_2$ \\
    p2mm& $3\mathbb Z_2\oplus\mathbb Z$ & $8\mathbb Z_2\rightarrow7\mathbb Z_2$\\
    p2mg& $2\mathbb Z_2\oplus\mathbb Z$ & $3\mathbb Z_2\rightarrow3\mathbb Z_2$\\
    p2gg& $\mathbb Z_2\oplus\mathbb Z$ & $2\mathbb Z_2\rightarrow2\mathbb Z_2$\\
    c2mm& $2\mathbb Z_2\oplus\mathbb Z$ & $5\mathbb Z_2\rightarrow4\mathbb Z_2$\\
    p4& $\mathbb Z_2\oplus\mathbb Z_4\oplus\mathbb Z$ & $\mathbb Z_2\oplus2\mathbb Z_4\rightarrow\mathbb Z_2\oplus2\mathbb Z_4$ \\
    p4mm& $\mathbb Z_2\oplus\mathbb Z_4\oplus\mathbb Z$ & $6\mathbb Z_2\rightarrow5\mathbb Z_2$\\
    p4gm& $\mathbb Z_4\oplus\mathbb Z$ & $2\mathbb Z_2\oplus\mathbb Z_4\rightarrow\mathbb Z_2\oplus\mathbb Z_4$\\
    p3& $2\mathbb Z_3\oplus\mathbb Z$ & $3\mathbb Z_3\rightarrow3\mathbb Z_3$\\
    p3m1& $2\mathbb Z_3\oplus\mathbb Z$ & $\mathbb Z_2\rightarrow\mathbb Z_2$\\
    p31m& $\mathbb Z_3\oplus\mathbb Z$ & $\mathbb Z_6\rightarrow\mathbb Z_6$\\
    p6& $\mathbb Z_6\oplus\mathbb Z$ & $2\mathbb Z_6\rightarrow2\mathbb Z_6$\\
    p6mm& $\mathbb Z_6\oplus\mathbb Z$& $4\mathbb Z_2\rightarrow3\mathbb Z_2$\\
    \hline\hline
  \end{tabularx}
\end{table}

\end{widetext}

\end{document}